\documentclass[manuscript,screen]{acmart}
\setcitestyle{sort&compress}
\usepackage{booktabs} 

\usepackage{amssymb}
\usepackage{amsmath}
\usepackage{amsfonts}
\usepackage{mathrsfs}
\usepackage{xspace}
\usepackage{srcltx}
\usepackage{caption}
\usepackage[labelformat=simple]{subcaption}
\usepackage{layouts}

\usepackage{pgfplots}

\pgfkeys{/pgf/number format/.cd,precision=2,1000 sep={}}

\usepackage[textwidth=1.2in]{todonotes}

\usepackage{listings}

\usepackage[linesnumbered,lined,commentsnumbered,ruled]{algorithm2e}

\newcommand*{\gz}[1]{\boldsymbol{#1}}
\renewcommand*{\d}{\mathrm{d}}

\usepackage{xparse}
\ExplSyntaxOn
\NewDocumentCommand{\mref}{m}{\quinn_mref:n {#1}}
\seq_new:N \l_quinn_mref_seq
\cs_new:Npn \quinn_mref:n #1
 {
  \seq_set_split:Nnn \l_quinn_mref_seq { , } { #1 }
  \seq_pop_right:NN \l_quinn_mref_seq \l_tmpa_tl
  ( 
  \seq_map_inline:Nn \l_quinn_mref_seq
    { \ref{##1},\nobreakspace } 
  \exp_args:NV \ref \l_tmpa_tl 
  ) 
 }
\ExplSyntaxOff

\acmJournal{TOPC}
\acmVolume{X}
\acmNumber{2}
\acmArticle{3}
\acmYear{2019}
\acmMonth{6}

\setcopyright{acmcopyright}

\acmDOI{0000001.0000001}

\def\dataptwoBandWidthHBcsr{7.3773993}

\def\dataptwoGFlopsHBcsr{24.0325348}
\def\dataptwoGFlopsHBcsrMFOnly{45.5392875}
\def\dataptwoGFlopsHBcsrNoRW{32.3306402}

\def\dataptwoStackIntegrals{47.2}
\def\dataptwoStackLoop{3.6}
\def\dataptwoStackMPI{21.3}
\def\dataptwoStackReadWrite{22.1}

\def\dataptwoWallHBcsr{0.58075}
\def\dataptwoWallHBcsrMFOnly{0.37519}

\def\dataptwoWallHFull{2.177}

\def\dataptwoWallMinHBcsrMFOnly{0.22254}

\def\dataptwoWallTmmultBcsr{0.19965}
\def\dataptwoWallTmmultFull{16.09665}
\def\dataptwoWallmmultBcsr{0.87748}
\def\dataptwoWallmmultFull{11.25986}
\def\dataptwobcsrmemory{607.788}
\def\dataptwocolblocksmean{8.42105}
\def\dataptwocolblocksmpimean{3.8}
\def\dataptwocolblocksmpistd{1.31656}
\def\dataptwocolblocksstd{1.15388}
\def\dataptwocolumns{320}

\def\dataptwodofs{846369}
\def\dataptwodofspercell{27}

\def\dataptwofullmemory{2292.35}
\def\dataptwomfnonemtpycolumnblocksmean{97429.8}
\def\dataptwomfnonemtpycolumnblocksstd{16796.0}
\def\dataptwoncells{102400}
\def\dataptwoncolblocks{38}
\def\dataptwonrowblocks{12800}
\def\dataptworowblocksmean{66.1226}
\def\dataptworowblocksstd{18.1344}

\def\datapfourGFlopsHBcsr{32.8855621}

\def\datapfourWallHBcsr{0.44389}

\def\datapfourWallHFull{1.36122}

\def\datapfourWallTmmultBcsr{0.44916}
\def\datapfourWallTmmultFull{16.01515}
\def\datapfourWallmmultBcsr{0.9374}
\def\datapfourWallmmultFull{12.33702}
\def\datapfourbcsrmemory{717.715}
\def\datapfourcolblocksmean{8.42105}
\def\datapfourcolblocksmpimean{3.8}
\def\datapfourcolblocksmpistd{1.31656}
\def\datapfourcolblocksstd{1.05604}
\def\datapfourcolumns{320}

\def\datapfourdofs{846369}
\def\datapfourdofspercell{125}

\def\datapfourmfnonemtpycolumnblocksmean{14877.4}
\def\datapfourmfnonemtpycolumnblocksstd{2685.25}
\def\datapfourncells{12800}
\def\datapfourncolblocks{38}
\def\datapfournrowblocks{1600}
\def\datapfourrowblocksmean{528.981}
\def\datapfourrowblocksstd{71.5597}

\def\dataGCCptwoBandWidthHBcsr{10.8588537}

\def\dataGCCptwoGFlopsHBcsr{34.5610683}

\def\dataGCCptwoStackIntegrals{42.6}
\def\dataGCCptwoStackLoop{4.8}

\def\dataGCCpfourGFlopsHBcsr{41.7628579}

\def\dataGCCpfourStackMPI{21.9}

\def\dataGCCpfourStackZero{14.1}
\def\dataGCCpfourWallHBcsr{0.34012}

\def\datacslptwoBandWidthHBcsr{38.6081869}

\def\datacslptwoGFlopsHBcsr{136.4514071}

\def\datacslptwocolblocksmean{9.27536}
\def\datacslptwocolblocksmpimean{3.45}
\def\datacslptwocolblocksmpistd{1.23438}
\def\datacslptwocolblocksstd{2.27444}
\def\datacslptwocolumns{640}

\def\datacslptwodofs{1686177}
\def\datacslptwodofspercell{27}

\def\datacslptwomfnonemtpycolumnblocksmean{100993.0}
\def\datacslptwomfnonemtpycolumnblocksstd{21082.6}
\def\datacslptwoncells{204800}
\def\datacslptwoncolblocks{69}
\def\datacslptwonrowblocks{25600}
\def\datacslptworowblocksmean{65.8663}
\def\datacslptworowblocksstd{18.2091}
\def\datacslpfourBandWidthHBcsr{58.0868785}

\def\datacslpfourGFlopsHBcsr{156.9741304}

\def\datacslpfourStackMPI{16.7}

\def\datacslpfourStackZero{26.9}
\def\datacslpfourWallHBcsr{0.21557}

\def\datacslpfourcolblocksmean{9.01408}
\def\datacslpfourcolblocksmpimean{3.55}
\def\datacslpfourcolblocksmpistd{1.31689}
\def\datacslpfourcolblocksstd{2.05282}
\def\datacslpfourcolumns{640}

\def\datacslpfourdofs{1686177}
\def\datacslpfourdofspercell{125}

\def\datacslpfourmfnonemtpycolumnblocksmean{15728.4}
\def\datacslpfourmfnonemtpycolumnblocksstd{3551.67}
\def\datacslpfourncells{25600}
\def\datacslpfourncolblocks{71}
\def\datacslpfournrowblocks{3200}
\def\datacslpfourrowblocksmean{526.93}
\def\datacslpfourrowblocksstd{71.2442}

\begin{document}

\title{%
Algorithms and data structures for matrix-free finite element operators with MPI-parallel sparse multi-vectors
}

\author{Denis Davydov}
\orcid{0000-0002-3779-0101}
\affiliation{
\institution{Chair of Applied Mechanics, Friedrich-Alexander-Universit\"{a}t Erlangen-N\"{u}rnberg}
\streetaddress{Egerlandstr.~5}
\city{Erlangen}
\postcode{91058}
\country{Germany}}
\email{denis.davydov@fau.de}

\author{Martin Kronbichler}
\orcid{0000-0001-8406-835X}
\affiliation{%
  \institution{Institute for Computational Mechanics, Technical University of Munich}
  \streetaddress{Boltzmannstr.~15}
  \city{Garching}
  \postcode{85748}
  \country{Germany}}
\email{kronbichler@lnm.mw.tum.de}

\renewcommand\shortauthors{Davydov, D. and Kronbichler, M.}

\begin{abstract}
Traditional solution approaches for problems in quantum mechanics scale as $\mathcal O(M^3)$, where $M$ is the number of electrons. Various methods have been proposed to address this issue and obtain linear scaling $\mathcal O(M)$. One promising formulation is the direct minimization of energy. Such methods take advantage of physical localization of the solution, namely that the solution can be sought in terms of non-orthogonal orbitals with local support.

In this work a numerically efficient implementation of sparse parallel vectors within the open-source finite element library \texttt{deal.II} is proposed. The main algorithmic ingredient is the matrix-free evaluation of the Hamiltonian operator by cell-wise quadrature. Based on an \textit{a-priori} chosen support for each vector we develop algorithms and data structures to perform (i) matrix-free sparse matrix multivector products (SpMM), (ii) the projection of an operator onto a sparse sub-space (inner products), and (iii) post-multiplication of a sparse multivector with a square matrix. The node-level performance is analyzed using a roofline model. Our matrix-free implementation of finite element operators with sparse multivectors achieves the performance of \pgfmathprintnumber[precision=1]{\datacslpfourGFlopsHBcsr} GFlop/s on Intel Cascade Lake architecture. Strong and weak scaling results are reported for a typical benchmark problem using quadratic and quartic finite element bases.
\end{abstract}


%
%


\maketitle

\section{Introduction}
\label{sec:intro}
In recent years there has been a growing interest in using finite element (FE) discretizations for problems in quantum mechanics
\cite{Pask2001,Tsuchida1996,Tsuchida1998,Fattebert:2007fy,Motamarri2012,Pask2017,RamMohan:2002ux,Davydov2017,Davydov2018,Bao2012,Fang2012,Young:2013cl,Zhang:2008dp,Bylaska:2009im,Schauer2014}, including an open-source production-ready FE-based implementation \cite{Motamarri2019} of the Density Functional Theory (DFT) \cite{Hohenberg:1964ut,Kohn:1965ui}.
An FE basis has the following advantages:
(i) it is a locally adaptive basis with variational convergence;
(ii) it has well-developed rigorous error estimates that can be used to efficiently drive locally adaptive refinement;
(iii) a hierarchy of nested functional spaces can be used to formulate geometric multigrid (GMG) preconditioners and solvers \cite{Davydov2018,Beck2009}, that are important
for the computationally efficient solution of source problems
and can also improve convergence of direct minimization methods \cite{Davydov2018,Fattebert2000};
(iv) an MPI-parallel implementation is achieved by
domain decomposition of the mesh and
does not require global communication;
(v) the FE formulation can be equipped with both periodic and non-periodic boundary conditions;
(vi) high order polynomial bases can be efficiently employed using matrix-free sum factorization approaches \cite{Kronbichler2012, Kronbichler2019};
(vii) pseudo-potentials and all electron calculations can be treated within the same framework.
Some of those advantages were employed to develop an open-source DFT-FE code \cite{Motamarri2019}.
The code was shown to outperform several commonly used plane-wave codes for systems with more than a few thousands of electrons.

Being a real-space method, an FE basis is also suitable for orbital minimization (OM) approaches
\cite{
    Hirose2001,Hirose2005, Mauri1994, Ordejon1995, Tsuchida1998, King-Smith1994,
    Peng2013,Burger2008,
    Fattebert2000, Fattebert2004, Galli1992,Fattebert:2007fy,
    Bowler2000a, Goringe1997,
    Davydov2018}
which can be combined with localization in real space to achieve linear scaling with respect to the number of unknown vectors $M$ in electronic structure calculations. In this case the solution is sought in terms of non-orthogonal orbitals with local support. Note that traditional eigensolver-based solution strategies have an $\mathcal O(M^3)$ computational complexity and an $\mathcal O(M^2)$ memory requirement.

Although finite elements have been used with localized orbitals \cite{Tsuchida1998}, to the best of our knowledge the implementation details of the underlying linear algebra have not been discussed. In the context of finite differences (FD), the authors of \citep{Fattebert2006} address this problem by ``packing together several nonoverlapping localized orbitals into a single global array that represents a grid-based function over the whole discretization grid'' using a graph-coloring algorithm.
However implementational details, performance and scaling studies were not presented.

In this contribution we propose algorithms and data structures for matrix-free finite element operators with MPI-parallel sparse multivectors within the \texttt{deal.II} finite element library \cite{dealII90}.
Implementational details are thoroughly described for three key operations:
(i) application of a matrix-free operator to a sparse multivector;
(ii) multivector inner products;
(iii) post-multiplication of a sparse multivector with a sparse square matrix.
Node level performance is analyzed using the roofline performance model.
Performance of the matrix-free operator is compared to the element stiffness-matrix based operators, adopted in \cite{Motamarri2019}.
Strong and weak scaling are studied on a typical example in $\mathbb R^3$.

\section{Required operations}
\label{sec:required_op}
Within the OM solution approach of DFT, the typical problem is to find the minimum of the functional
\begin{align}
E(\gz \Phi) = {\rm tr}(
    [
        2\overline{\gz I} -
        \underbrace{\gz \Phi^T \gz M \gz \Phi}_{\displaystyle\overline{\gz{M}}}
    ]
    \underbrace{\gz \Phi^T \gz H \gz \Phi}_{\displaystyle\overline{\gz H}}
    ) \, .
\label{eq:energy}
\end{align}
Here $\gz \Phi = \{ \gz \phi_1, \gz \phi_2, \dots , \gz \phi_M  \} \in \mathbb R^{N\times M}$ denote a set of $M$ vectors (a ``multivector'') where each vector is of size $N$.
$\gz M \in \mathbb R^{N\times N} $ and $\gz H \in \mathbb R^{N\times N}$ are symmetric matrices representing the mass and Hamiltonian operators in the FE basis.
For the current study we limit ourselves to a local potential. In this case the Hamiltonian operator has the form $ -\frac{1}{2} \Delta + V(\gz x)$, where $V(\gz x)$ is a given scalar-valued potential field.
Given that FE shape functions have local support, both $\gz M$ and $\gz H$ are sparse.
$\overline{\gz H}\in \mathbb R^{M\times M}$ and $\overline{\gz M}\in \mathbb R^{M\times M}$ are projections of mass and Hamiltonian operators onto the subspace spanned by $\gz \Phi$.
The number of columns $M$ represents the number of unknown vectors (on the order of the total number of electrons in the systems), whereas the number of rows $N$ is related to the spatial discretization using the FE basis.
Within the FE context applied to such problems, the multivector $\gz \Phi$ is tall and skinny, i.e., $N \gg M$.
Here and below we denote by $\overline{\{\bullet \}}$ comparatively small matrices that do not contain $N$ as either of their dimensions.

The gradient of \eqref{eq:energy} reads
\begin{align}
\gz G(\gz \Phi) = -2 \gz M \gz \Phi \overline{\gz H} +
                   2 \gz H \gz \Phi [ 2 \overline{\gz I} - \overline{\gz M}] \, .
\label{eq:gradient}
\end{align}
Additionally the directional derivative of the functional \eqref{eq:energy} is required for line searches and is given by the sum of inner products for each column $\alpha$,
\begin{align}
    \frac{\d E(\gz \Phi + \epsilon \gz D)}{\d \epsilon} \Bigr\rvert_{\epsilon=0} =
    {\rm{tr}}(\gz D^T \gz G) = \sum_{\alpha} \gz D_{\alpha} \cdot \gz G_{\alpha} \, .
\end{align}

In order to perform direct minimization (i.e., by adopting the steepest descent method or BFGS \cite{Davydov2018}), the following operations are required:
\begin{itemize}
    \item sparse-matrix multivector multiplications:
    \begin{align}
        \gz \Phi_M = \gz M \gz \Phi\, , \\
        \gz \Phi_H = \gz H \gz \Phi\, ,
    \end{align}
    \item multivector inner products:
    \begin{align}
        \overline{\gz M} &= \gz \Phi^T_M \gz \Phi=\gz \Phi^T \gz \Phi_M\, , \label{eq:inner1} \\
        \overline{\gz H} &= \gz \Phi^T_H \gz \Phi=\gz \Phi^T \gz \Phi_H \, , \label{eq:inner2}
    \end{align}
    where $\overline{\gz M}$ and $\overline{\gz H}$ are symmetric Hermitian (thanks to the properties of $\gz M$ and $\gz H$).
    \item right-multiplication with a square matrix: 
    \begin{align}
        \gz G &= -2 \gz \Phi_M \overline{\gz H}\, , \label{eq:G_ops1} \\
        \gz G &=  2 \gz \Phi_H [ 2 \overline{\gz I} - \overline{\gz M}] + \gz G\, , \label{eq:G_ops2}
    \end{align}
    where $\gz G \in \mathbb R^{N\times M}$.
    \item column-wise inner product:
    \begin{align}
        g = {\rm{tr}}(\gz D^T \gz G) = \sum_{\alpha} \gz D_{\alpha} \cdot \gz G_{\alpha}\, ,
        \label{eq:G_ops4}
    \end{align}
    where $\gz D \in \mathbb R^{N\times M}$ is the search direction.
\end{itemize}

Our goal is to implement these operations for the case when the minimum of \eqref{eq:energy} is sought in terms of an \textit{a-priori}
chosen\footnote{An alternative approach is to augment \eqref{eq:energy} to favor sparsity (e.g.~using the $l^1$-regularization \cite{Lu2017})}
sparsity of an unknown multivector $\gz \Phi$.
Note that in this case $\overline{\gz H}$ and $\overline{\gz M}$ are also sparse.
We are interested in the case when the number of unknown multivectors $M$
is on the order of a few thousands, whereas the number of basis functions $N$ in the FE basis is several orders of magnitude larger.
Consequently, the operations we focus on are those that contain the largest dimension $N$, namely application of mass and Hamiltonian operators, multivector inner products and post-multiplication.

\subsection{Available parallel sparse linear algebra packages}
Given the sparse structure of $\gz \Phi$, the operations listed above are essentially sparse matrix-matrix multiplications (SpMM).
Below we give a brief overview of existing open source libraries focusing on the required operations and explain why they are not well-suited for our application.

GHOST \cite{Kreutzer2017} is a relatively new high-performance sparse linear algebra package aimed at heterogeneous systems. Sparse matrices are stored in the SELL-C-$\sigma$ format that allows to employ SIMD vectorization over a chunk of rows in a matrix and achieve very competitive implementation of sparse matrix vector (SpMV) products \cite{Kreutzer2014}. The storage format, however, is not well suited for random access to rows of the matrix, which is the case of matrix-free FE operators applied to sparse multivectors (that we shall discuss in the next sections).
Also, at the time of writing, GHOST does not provide any kernels for SpMM multiplication, the \texttt{ghost\_tsmttsm} kernel only supports dense tall and skinny matrices.

A commonly used linear algebra package within the FE community is \texttt{TPetra}\footnote{We do not consider \texttt{EPetra} as all ongoing efforts on linear algebra packages within Trilinos are concentrated on \texttt{TPetra}.} \cite{Nusbaum2011} from the Trilinos \cite{Heroux2005} suite.
It provides sparse matrices in either compressed row storage (CRS) or compressed column storage (CCS).
However, sparse matrix-matrix multiplication kernels provided by \texttt{Tpetra::MatrixMatrix::Multiply} are not flexible enough for our needs.
More precisely, when SpMM is to be performed multiple times for the same structure of matrices, the sparsity of the destination matrix should be consistent to the product.
For the case of non-local potentials, this will not hold for \mref{eq:G_ops1,eq:G_ops2}, where each of those products will result in different sparsity patterns and the sparsity pattern of $\gz G$ is their union.
SpMM in \texttt{TPetra} also does not take into account the symmetry of \mref{eq:inner1,eq:inner2}, originating from the symmetry of mass and Hamiltonian operators.
Additionally, we mention that $\gz A^T \gz B$ creates an actual transpose of the matrix prior to evaluation of the product \cite{Nusbaum2011}.
Finally, within the matrix-free context of FE operators source (multi)vectors are required to keep information not only about locally owned rows (assuming 1D row-wise partitioning of $\gz \Phi$), but also know values for a relatively small number of rows owned by other MPI processes. Within the \texttt{TPetra} implementation model that would require two instances of $\gz \Phi$, which would effectively double the memory footprint.

Another commonly used alternative is the PETSc \cite{petsc-user-ref} package. Similar limitations regarding the sparsity of the destination matrix, evaluation of transpose and symmetry also apply to PETSc's \texttt{MatTransposeMatMult} implementation of SpMM multiplication kernel.
Additionally, at the time of writing, PETSc does not support shared memory parallelization.

Finally, the sparsity structure of $\gz \Phi$ is relatively dense compared to the sparsities originating from FEM or FD discretizations
of local operators in PDEs, which are one of the main application scenarios for the packages discussed above.

\section{Implementation details}
\label{sec:implementation}
In this section we detail the main algorithmic building blocks for high-performance DFT calculations with finite elements using sparse vectors.

\subsection{Finite element method and MPI parallelization}
\label{sec:parallel_FE_vectors}

\begin{figure}
    \begin{subfigure}[b]{0.32\textwidth}
        \centering
        \includegraphics[width=\textwidth]{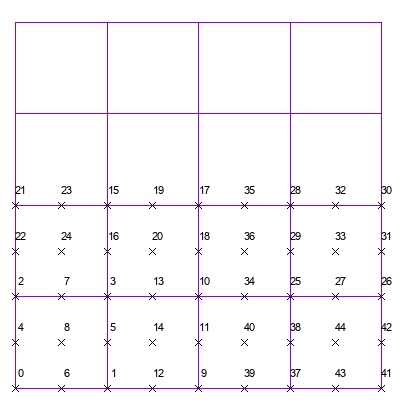}
        \caption{Process 0, owns $[0,44]$}
        \label{fig:owned_proc0}
    \end{subfigure}
    ~
    \begin{subfigure}[b]{0.32\textwidth}
        \centering
        \includegraphics[width=\textwidth]{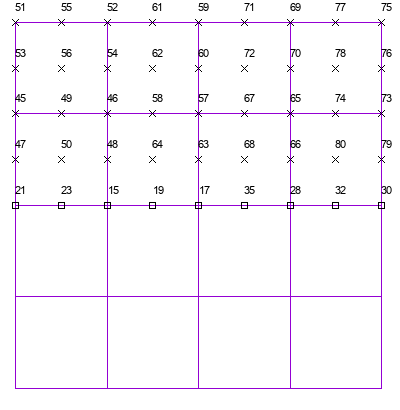}
        \caption{Process 1, owns $[45,80]$}
        \label{fig:owned_proc1}
    \end{subfigure}
    ~
    \begin{subfigure}[b]{0.32\textwidth}
        \centering
        \includegraphics[width=\textwidth]{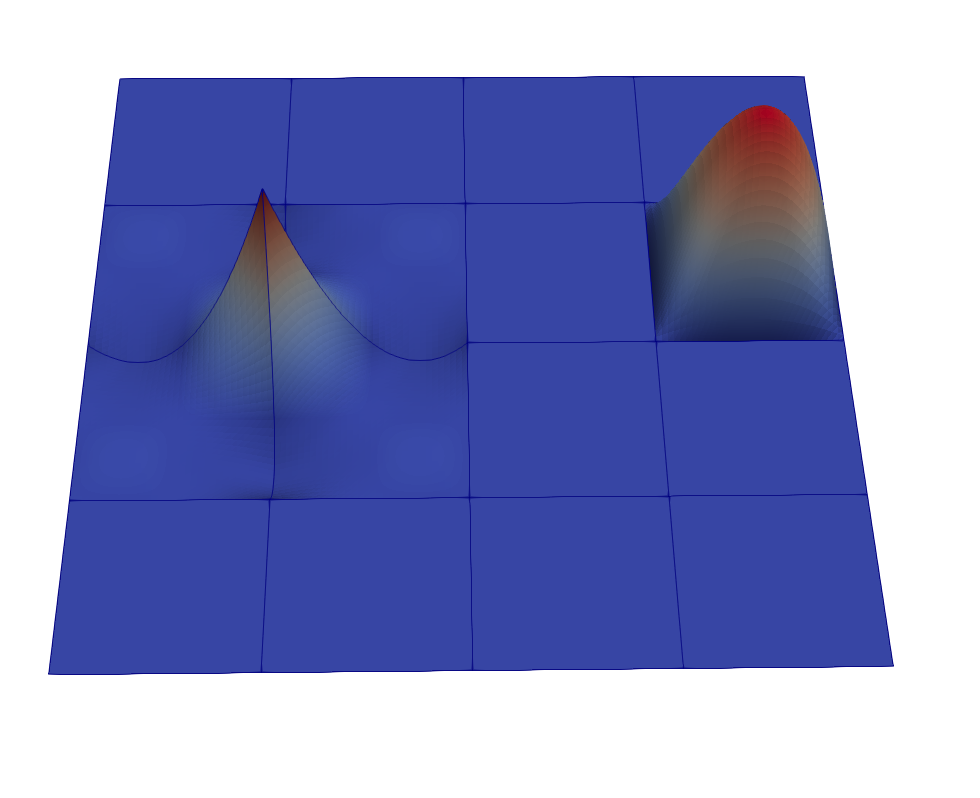}
        \caption{basis functions $N_{23}(\gz x)$ and $N_{80}(\gz x)$}
        \label{fig:fe_shape_func}
    \end{subfigure}
    \caption{FE mesh with quadratic Lagrange basis partitioned into 2 processes. Locally owned DoFs are depicted with crosses whereas ghost DoFs are depicted with squares.}
    \label{fig:fe_vectors}
\end{figure}

We start by summarizing the MPI parallelization of the finite element method with standard dense vectors and related concepts needed for matrix-free operator evaluation \cite{Kronbichler2012}.
First we introduce a FE triangulation $\mathcal{T}^h$ of the domain $\Omega$ and an associated
FE space of continuous piecewise shape functions of a fixed polynomial degree $p$ in $d$ spatial dimension $\phi_\alpha^h \in V^h \subset H^1_0 (\Omega)$.
The unknown orbital fields (a multivector) are given in a vector space spanned by standard FE basis functions $\{N_i(\gz x)\}$ (e.g. polynomials with local support):
\begin{align}
\phi_\alpha^h (\gz x) &=: \Phi_{i \alpha} N_i(\gz x)\, ,
\end{align}
where the superscript $h$ indicates that this representation is related to the FE mesh with size function $h(\gz x)$. In this work, we consider shape functions that are Lagrange polynomials on the elements and continuous over element boundaries. The interpolant $\phi_\alpha^h$ can be described in terms of its values $\Phi_{i \alpha}$ in the points $\gz x_i$ via the interpolation property $N_j(\gz x_i) = \delta_{ij}$. These interpolation points are called ``nodes'' in the following.
It is common within the FE community to refer to node values as degrees of freedom (DoFs) since the discretization
is typically applied to source problems, in which case the number of unknowns is the same as the number of basis functions.
Below we follow the same notation and effectively denote by DoFs the rows in the unknown multivector $\gz \Phi$.

The entries of the mass and Hamiltonian matrices are given by
\begin{equation}
\begin{split}
M_{ij} &= \sum_{K} \int_{\Omega^h_{K}} N_i(\gz x) N_j(\gz x) d\gz x \, , \\
H_{ij} &= \sum_{K} \int_{\Omega^h_{K}} \left[\frac{1}{2} \gz \nabla N_i(\gz x) \cdot \gz \nabla N_j(\gz x) + N_i(\gz x) V(\gz x) N_j(\gz x)\right] d\gz x \, ,
\end{split}
\label{eq:fe_matrices}
\end{equation}
where $\Omega^h_{K}$ is an element $K$ of the triangulation 
of the domain $\Omega^h = \bigcup_K \Omega^h_{K}$. In this work, we consider hexahedral elements.

As an example consider a setup with quadratic elements on a mesh partitioned into 2 MPI processes, see Figure \ref{fig:fe_vectors}.
Elements are partitioned one-to-one into a given number of MPI processes (e.g., using graph coloring or space-filling curves).
Based on the mesh partitioning, index sets of locally ``owned'' DoFs is constructed.
Note that the relationship between the ``owned'' DoFs and MPI processes is a one-to-one map.
In Figure \ref{fig:owned_proc0} and \ref{fig:owned_proc1} such DoFs are marked with crosses.
Clearly DoFs that correspond to nodes at the interface between two MPI domains can be assigned to either process.
In \texttt{deal.II} the process with a lower rank will claim the ownership.
Figure \ref{fig:fe_shape_func} depicts two shape functions for the given triangulation and a quadratic Lagrange basis.
Given the local support of the shape functions $N_i(\gz x)$, the matrices $M_{ij}$ and $H_{ij}$ are sparse,
with a sparsity pattern implied by the element connectivity and the adopted basis.
A common realization of FE algorithms is to numerically evaluate the integrals in \eqref{eq:fe_matrices} and store non-zero elements in a sparse matrix organized in e.g.~the compressed sparse row (CSR) format.
Sparse matrix-vector products can then be evaluated in parallel based on a one-to-one partitioning of DoFs and matrix rows.

\subsection{Matrix-free approach}
\label{sec:matrix_free}

\begin{figure}
    \centering
    \includegraphics[width=0.5\textwidth]{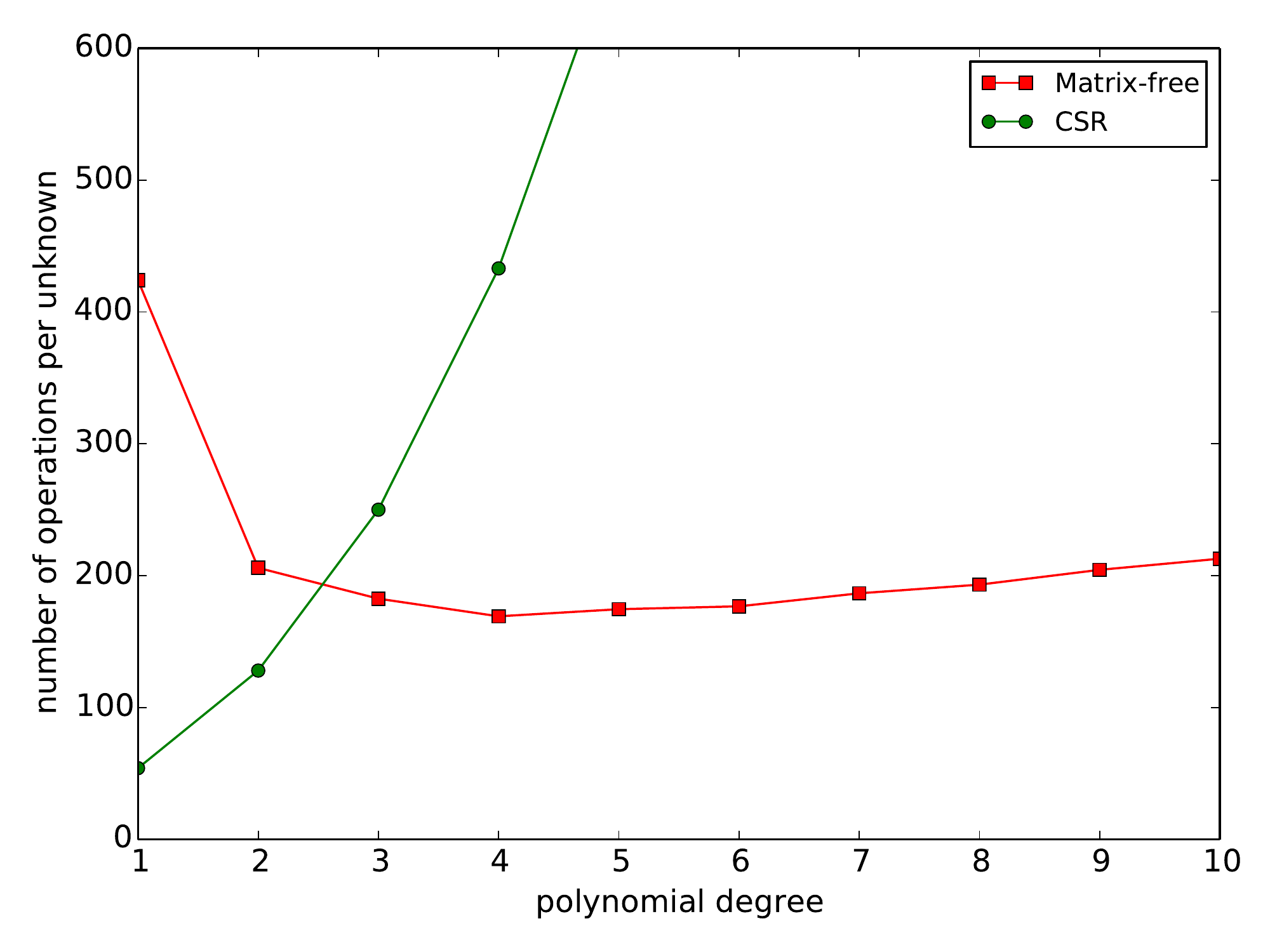}
    \caption{Arithmetic complexity of matrix-free and matrix-based CSR approaches for a single matrix-vector multiplication with the Hamiltonian operator for cartesian mesh with periodic boundary conditions.}
    \label{fig:complexity}
\end{figure}

As an alternative to the global assembled view of FE matrices with global DoF numbers $i$ and $j$, one can take an element-based view of the operators,
\begin{equation}
    \gz M = \sum_{K} \gz P_K^T \gz M_K \gz P_K\, , \qquad    \gz H = \sum_{K} \gz P_K^T \gz H_K \gz P_K\, .
    \label{eq:fe_matrices_cell_based}
\end{equation}
Here $\gz H_K \in \mathbb R^{[p+1]^d\times [p+1]^d}$ and $\gz M_K \in \mathbb R^{[p+1]^d\times [p+1]^d}$ are the element mass and Hamiltonian matrices,
which represent integrals of the basis functions inserted into the bilinear forms integrating over each element $K$,
and $p$ is the polynomial degree of the shape functions.
$\gz P_K$ is a sparse matrix which extracts local values on element $K$, whereas $\gz P_K^T$ is its inverse (i.e. $\gz P_K^T \gz P = \gz I$) that maps DoFs on a cell $K$ to global DoFs, typically implemented by an index array and indirect addressing. Here
and below we assume absence of constraints (e.g., due to hanging nodes). This case is not persued here, but it can be handled straight-forwardly by adopting algorithms and data structures developed for standard SpMV multiplication in \cite{Kronbichler2012}.

This view of the problem allows to evaluate the matrix-vector product, i.e., the discretized differential operator in terms of the finite element shape functions, in a matrix-free fashion \cite{Kronbichler2012, Kronbichler2019}.
Entries of the sparse matrices are neither explicitly calculated nor stored, and only the operator action on a (multi-)vector is requested in (iterative) solution algorithm. The local multiplications by $\gz M_K$ and $\gz H_K$ on local (multi-)vector are computed by a fast integration using sum factorization.
Sum factorization utilizes the tensor product structure of shape functions and quadrature points on hexahedra and expresses interpolations and derivatives in reference coordinate by a series of one-dimensional operations. The theoretical complexity of operator evaluation in terms of the polynomial degree $p$ is $\mathcal O(p)$ per DoF. The runtime often behaves as $\mathcal O(1)$ because memory bandwidth is the limiting factor rather than arithmetic. Matrix-based CSR approaches lose the tensor-product structure as soon as non-affine meshes are considered or operators with non-constant coefficients are involved, and give rise to arithmetic and memory complexities of $\mathcal O(p^3)$ in three space dimensions.
The arithmetic work, expressed as the number of floating point operations per DoF, to evaluate the Hamiltonian operator with a matrix-free approach using fast integration is compared to a matrix-based CSR approach in Figure \ref{fig:complexity} using data from \cite{Kronbichler2019}.
Even though computing integrals on the fly clearly increases work compared to the final sparse matrix for $p=1$, sum factorization eventually reverses the picture. Note that the initial reduction of the work per unknown in the matrix-free case is due to the reduced proportion of DoFs shared by several elements, that contribute with separate arithmetic operations on each element.

The matrix-free method also comes along with a significantly reduced memory consumption. In classical finite element implementations where the operator is applied to a single vector, the lower memory transfer implies that the case $p=2$ is up to five times faster than the sparse matrix-vector product, despite a higher arithmetic work \cite{Kronbichler2019}, given the fact that SpMV is memory-bandwidth limited on all current hardware architectures. In the context of the DFT, the operator is applied to a multivector and the matrix entries can be reused for several vectors. SpMMV implementations with a considerably higher utilization of floating point units have been developed, see e.g.~the recent work by Anzt et al.~\cite{Anzt2015} and Hong et al.~\cite{Hong2018} for GPUs. An alternative approach adopted in the DFT context for SpMMV FE operators is to store element matrices $\gz M_K$ and $\gz H_K$ and resort to dense matrix-matrix products on the element level \cite{Motamarri2019}. In this work, we consider methods with moderate polynomial degrees $p=2$ and $p=4$ which balance the needs for localization (favoring many low-degree elements) and better accuracy per DoF of higher-degree elements in terms of the underlying orbitals. According to Figure~\ref{fig:complexity}, matrix-free approaches are the most favorable setup in this case with around 170 Flop/DoF.

The element-based matrix-free approach requires knowledge of the vector values associated with the locally owned elements.
At the interface between MPI domains this requires access to so-called ``ghost'' DoFs, that are owned by another MPI processes (depicted by squares in Figure \ref{fig:owned_proc1}).
Within the \texttt{deal.II} library, the required MPI communication is termed ``update ghost values'' .
This is a synchronization operation which copies some values from process which ``owns'' a DoF to all processes which access that DoF as a ``ghost'' DoF, ensuring access to the replicated data.

On the other hand, the loop over cells according to \eqref{eq:fe_matrices_cell_based} implies that the operator on cell $K$ will contribute to all DoFs on this cell in the destination (multi-)vector.
This necessitates a second MPI communication operation upon the completion of the cell loop in \eqref{eq:fe_matrices_cell_based} which sends accumulated data (representing partial integrals with the local variant of the test function) to the owning MPI process, which then adds it to the local integral contribution. We refer to that operation as ``compress''.

In order to ensure a direct array access for reading and writing to each locally owned DoF, MPI-parallel vectors in \texttt{deal.II} perform all index access using a process-local enumeration of DoFs, including first all locally owned DoFs and then ``ghost'' DoFs. The latter are grouped according to the rank of the MPI process owning them so that the non-blocking ``MPI\_Irecv'' in ``update ghost values'' and ``MPI\_Isend'' in ``compress'' can use this memory directly.
Note that an additional temporary array is needed to receive contributions to the locally owned DoFs during ``compress'' operation or send them during ``update ghost values'' operation in an unpack/pack fashion. This way, memory for the vector can be allocated as a single large block. Global operations on the vectors then only operate on a vector view of the locally owned DoFs, whereas cell loops operate on a view including the owned and ghost DoFs, without a deep copy between the two states.  By contrast, in the \texttt{TPetra} linear algebra one would need to create a fully distributed vector with one-to-one correspondence between DoFs and MPI ranks, and a vector with one-to-many relationship between DoFs and MPI ranks.
The operation labeled ``compress'' corresponds to the ``export'' operation from the vector with one-to-many map to vector with one-to-one map, whereas
the ``update ghost values'' corresponds to the ``import'' operation in the other direction. This ``view'' scenario obviates allocating two vectors and will be especially important for the sparse multivectors considered below. In order to avoid ambiguity in terms of the data in the ghosted arrays, the MPI-parallel vectors natively provided by \texttt{deal.II} are either in the state where it is allowed to write into the ``ghost'' DoFs followed by the ``compress'' operation, or when ``update ghost values'' is called and ghost DoFs contain values consistent with the owning process.
In the latter case only read operation is allowed.

\subsection{Sparse multivectors}
\label{sec:parallel_sparse_FE_vectors}

%
%
\begin{figure}[!ht]
    \hfill
    \begin{subfigure}[b]{0.32\textwidth}
        \centering
        \includegraphics[width=\textwidth]{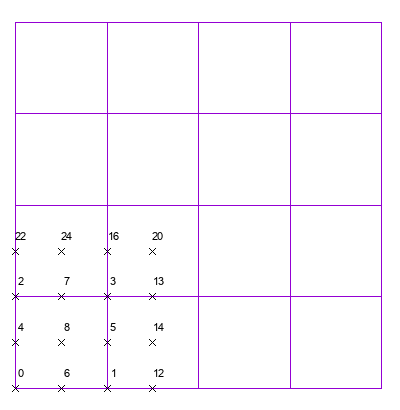}
        \caption{Process 0, support 0}
        \label{fig:owned_proc0_sparse0}
    \end{subfigure}
    ~
    \begin{subfigure}[b]{0.32\textwidth}
        \centering
        \includegraphics[width=\textwidth]{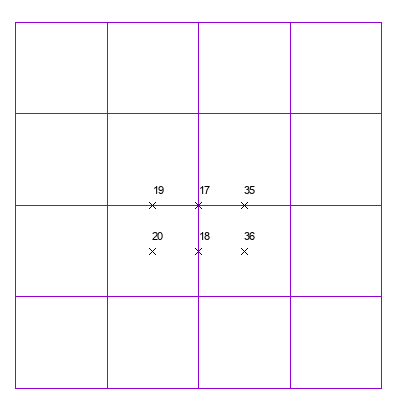}
        \caption{Process 0, support 1}
        \label{fig:owned_proc0_sparse1}
    \end{subfigure}
    ~
    \begin{subfigure}[b]{0.32\textwidth}
        \centering
        \includegraphics[width=\textwidth]{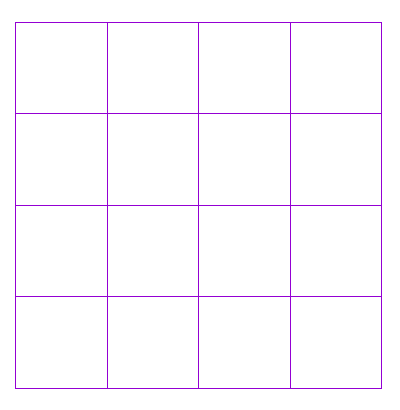}
        \caption{Process 0, support 2}
        \label{fig:owned_proc0_sparse2}
    \end{subfigure}
    \hfill
    \\
    \hfill
    \begin{subfigure}[b]{0.32\textwidth}
        \centering
        \includegraphics[width=\textwidth]{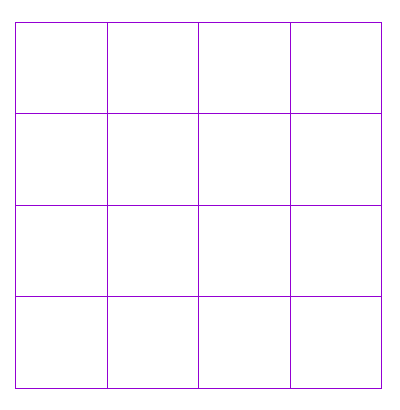}
        \caption{Process 1, support 0}
        \label{fig:owned_proc1_sparse0}
    \end{subfigure}
    ~
    \begin{subfigure}[b]{0.32\textwidth}
        \centering
        \includegraphics[width=\textwidth]{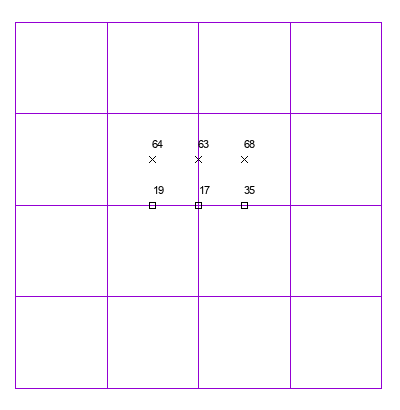}
        \caption{Process 1, support 1}
        \label{fig:owned_proc1_sparse1}
    \end{subfigure}
    ~
    \begin{subfigure}[b]{0.32\textwidth}
        \centering
        \includegraphics[width=\textwidth]{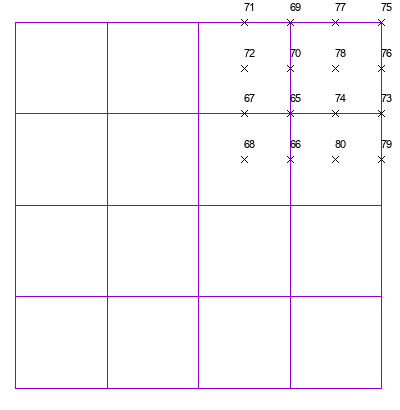}
        \caption{Process 1, support 2}
        \label{fig:owned_proc1_sparse2}
    \end{subfigure}
    \hfill
    \caption{Three support domains for sparse FE vectors distributed over two MPI processes.
    Non-zero locally owned DoFs are depicted with crosses whereas non-zero ghost DoFs are depicted with squares.}
    \label{fig:fe_vectors_sparse}
\end{figure}

Let us now introduce sparse FE multivectors.
Consider a set of overlapping subdomains $\{ \Omega_\alpha^h \}$ covering the domain $\Omega^h$ that satisfy the point-wise overlap condition
\begin{equation}
    \exists M_s \in \mathbb N \quad \forall \gz x \in \Omega^h \quad \rm{card} \{ \alpha\, | \, \gz x \in \Omega_\alpha^h  \} \leq M_s \, .
\end{equation}
In other words, not more than $M_s$ patches overlap at any point in the domain.
We are interested in the case when
\begin{equation}
    M_s \ll M\, ,
    \label{eq:overlap_condition}
\end{equation}
where each sparse FE vector is typically non-zero only on a handful of MPI processes irrespective of the number $P$ of MPI procsses that the domain $\Omega^h$ is partitioned into.

As an example, consider three different support domains $\Omega_\alpha^h \in \Omega^h$ for such vectors as illustrated in Figure \ref{fig:fe_vectors_sparse}.
We assume \textit{a-priori} knowledge about the sparsity of the FE multivectors, which is usually based on proximity to a given set of points in space (e.g., location of atoms).
In such applications it is either assumed or it can be shown that the solution can be represented with functions localized in space with exponential decay \cite{Ismail-Beigi1999,Kohn1959}.

Our implementation of sparse FE vectors is based on the blocked compressed sparse row (BCSR) format with 1D row-wise MPI partitioning.
BCSR has been adopted for problems in quantum mechanics \cite{Challacombe2000,Saravanan2003,Borstnik2014} with Gaussian bases.
BCSR matrix can be considered as an extension of CSR matrices where each element is a dense matrix.
Their sizes are specified by (independent) row and column blocking.
As a result, operations like SpMM products will eventually perform dense matrix-matrix multiplications for non-zero blocks,
typically using BLAS \texttt{dgemm} routines.
For matrices with a large number of non-zeroes, such a storage format is expected to be more efficient and have a higher performance.

An important step of BCSR is the re-ordering of rows (DoFs in the FEM context) and columns (unknown vectors) of sparse matrices.
To that end, Hilbert space filling curves (HSFC) are often adopted \cite{Challacombe2000, Saravanan2003}, which can be considered as a heuristic solution to the travelling salesman problem.

\begin{figure}[!ht]
    \begin{minipage}{0.49\textwidth}
    \centering
    \begin{subfigure}[b]{0.64\textwidth}
        \centering
        \includegraphics[width=\textwidth]{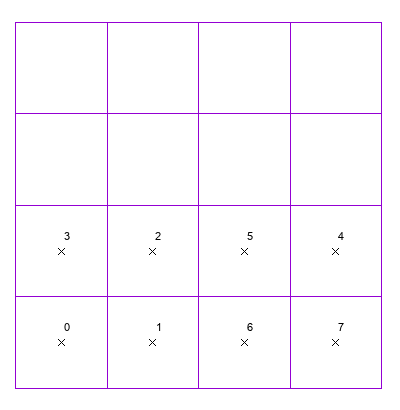}
        \caption{Cell order on process 0}
        \label{fig:cell_order_0}
    \end{subfigure}
    \\
    \begin{subfigure}[b]{0.64\textwidth}
        \centering
        \includegraphics[width=\textwidth]{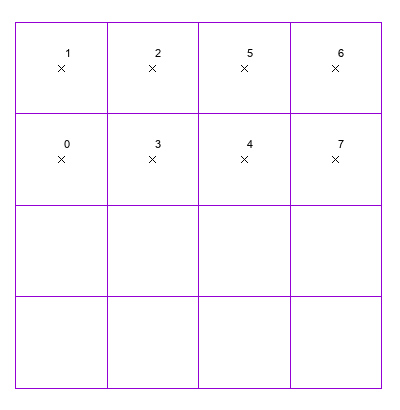}
        \caption{Cell order on process 1}
        \label{fig:cell_order_1}
    \end{subfigure}
    \end{minipage}
    \begin{minipage}{0.49\textwidth}
    \centering
    \begin{subfigure}[b]{0.64\textwidth}
        \centering
        \includegraphics[width=\textwidth]{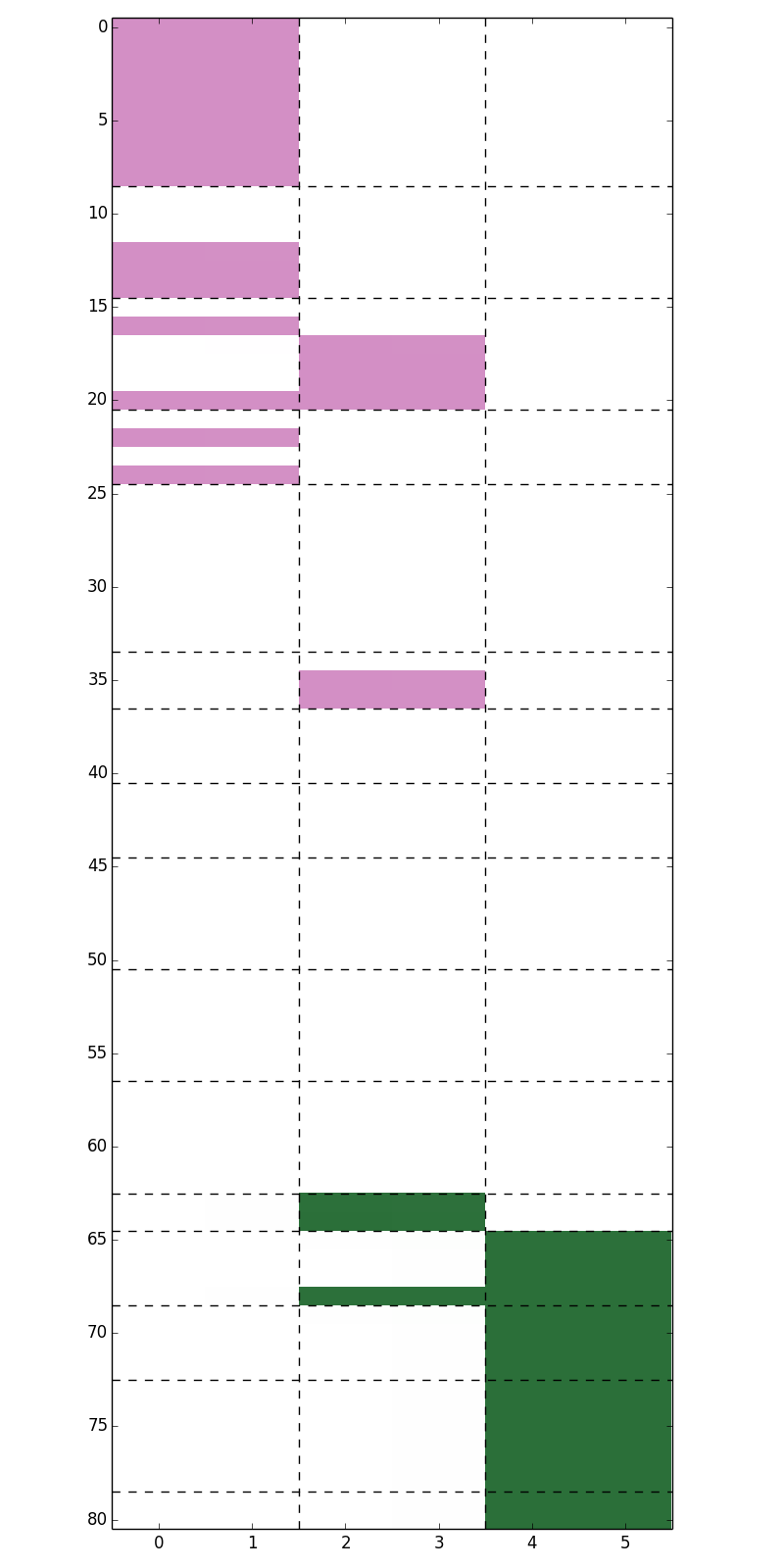}
        \caption{BCSR sparsity and blocking}
        \label{fig:bcsr_sparsity}
    \end{subfigure}
    \end{minipage}
    \caption{Renumbering of DoFs using Hilbert space filling curves via Algorithm \ref{alg:dof_renumbering}. Subfigures (a) and (b) depict order of locally owned cells.
    Subfigure (c) shows the BCSR representation of the six sparse vectors with support depicted in Figure \ref{fig:fe_vectors_sparse} (two sparse vectors per localization domain).
    Filled rectangles denote the non-zero ranges of those vectors, whereas their colors indicate the MPI process id that own the respective rows.
    Clearly BCSR matrices will store (slightly) more non-zeroes than required as illustrated by the not completely filled dense blocks in subfigure (c).
    }
    \label{fig:rows_renumbering_and_bcsr}
\end{figure}

\begin{algorithm}[h]
    \SetKwInOut{Input}{Given}
    \SetKwInOut{Output}{Return}
    \Input{a FE triangulatio partitioned into MPI domains, coarse level $L$.}
    \Output{renumbered DoFs, ordered active cells list, block sizes for rows}
    collect centers of cells on the chosen mesh level $L$ \;
    order coarse level cells using HSFC \;
    \ForEach{ordered coarse cell $K_L$} {
        traverse its children and add all active cells to the ordered list \;
        record the number of active cell $N_K$ added from this coarse cell \;
    }
    renumber locally owned DoFs based on the constructed order of active cells \;
    traverse the ordered list of active cells and create DoF blocks based on $\{N_K\}$ \;
    \caption{DoFs renumbering and blocking using Hilbert space filling curves.}
    \label{alg:dof_renumbering}
\end{algorithm}

In order to perform the renumbering of locally owned DoFs we propose Algorithm \ref{alg:dof_renumbering}.
An illustrative example of this algorithm is given in Figures \ref{fig:cell_order_0} and \ref{fig:cell_order_1}, producing a DoF numbering adopted in
Figure \ref{fig:fe_vectors_sparse}.
We emphasize that HSFC are not applied on the finesh mesh level (which could be constructed from adaptive mesh refinement (AMR)),
but possibly on a coarser level $L$. This gives additional control over the blocking of rows and combines well with
algorithms that construct support domains for sparse multivectors also by starting from some coarse level $L$. Although AMR is not considered in this manuscript,
we are convinced that support domains (in terms of element patches)
for sparse multivectors should be constructed based on a mesh without hanging nodes.
In this case, AMR will lead to geometrically equivalent support (i.e., a patch of elements will cover exactly the same domain)
and thus the variational property of the FE solution will be preserved.
The enumeration of DoFs based on the cell order (see step 7 in Algorithm \ref{alg:dof_renumbering}) is done using \texttt{DoFRenumbering::cell\_wise} function of \texttt{deal.II},
which assigns new DoF indices based on the ordering of cells. DoFs are assigned by going through the vertex, line, surface, and volume DoFs of each cell.
DoFs associated to several cells are numbered on the cell where they are encountered first.
Based on this idea, we can construct a blocking of DoFs using the ordered cells as well.
That is, each block consists of all DoFs in a group of active cells $N_K$ (see step 5 in Algorithm \ref{alg:dof_renumbering}) that have not yet
been assigned to other blocks.
The resulting row blocking is shown in Figure \ref{fig:bcsr_sparsity}.

\begin{algorithm}[h]
    \SetKwInOut{Input}{Given}
    \SetKwInOut{Output}{Return}
    \Input{a FE triangulation $\mathcal{T}^h$, partitioned into MPI domains, localization centers of sparse vectors $\{ \gz L_{\alpha} \}$, block size $B$}
    \Output{renumbered sparse vectors, block sizes for columns}
    for each center of the localization domain $\gz L_{\alpha}$ find a cell $K_\alpha$ it belongs to \;
    sort vectors according to the MPI rank of the owning process \;
    sort each group of vectors that belong to the same MPI process using HSFC \;
    within each MPI process, group vectors into blocks according to the specified blocking $B$ with remainder (if any) assigned to the last block \;
    \caption{Sparse vectors renumbering and blocking using Hilbert space filling curves.}
    \label{alg:vec_renumbering}
\end{algorithm}

Renumbering of vectors (columns) is also done using HSFC according to Algorithm \ref{alg:vec_renumbering}.
First we assign an MPI process to each column based on the location of the center of the localization domain within the
MPI-partitioned mesh. Given
that the MPI partitioning of the mesh is less likely to produce disjoint subdomains,
this distribution approach is similar to the one adopted in \cite{Bowler2001},
where several spatially compact groups of atoms are assigned to a single MPI process.
In order to speed up the search for a cell that contains the localization center (step 1 in Algorithm \ref{alg:vec_renumbering}),
we precompute the bounding box of all locally owned cells. This allows to skip the search for points $\gz L_\alpha$ that are outside of this box.
In order to further speed up the search for a cell, an rtree provided by \texttt{boost::index::rtree} is employed through the \texttt{GridTools::Cache} class of \texttt{deal.II}.
For $\gz L_\alpha$ that are located exactly at the interface between MPI domains of the mesh, we may find $K_\alpha$ it belongs to on different MPI processes.
To circumvent this ambiguity we assign the process with the lower MPI rank as the owner.
Note that the blocking of columns is typically physically motivated (e.g., orbitals belonging to the same atom can be grouped together).
For the current study and the numerical examples below, we perform the blocking based on a pre-defined size $B$ (step 4 in Algorithm \ref{alg:vec_renumbering}), but the algorithm is flexible in terms of the blocking parameters.
Algorithm \ref{alg:vec_renumbering} implies that the MPI partitioning of the triangulation together with the location of the localization centers completely determine the MPI partitioning of the columns in the multivector.
Note that the MPI partitioning of the solution vectors is not affected and rather only the projected matrices $\overline{\gz M}$ and $\overline{\gz H}$.
In the case of a large number of MPI processes and relatively few columns,
some processes may not ``own'' any localization domains,
but they may still contribute to projected matrices, unless a locally owned subdomain does
not overlap with any support of sparse multivectors.
An illustration of the column enumeration is given in Figure \ref{fig:bcsr_sparsity},
where we assume two sparse vectors per localization domain from Figure \ref{fig:fe_vectors_sparse}.

\begin{figure}
    \centering
    \begin{subfigure}[c]{0.32\textwidth}
        \includegraphics[width=\textwidth]{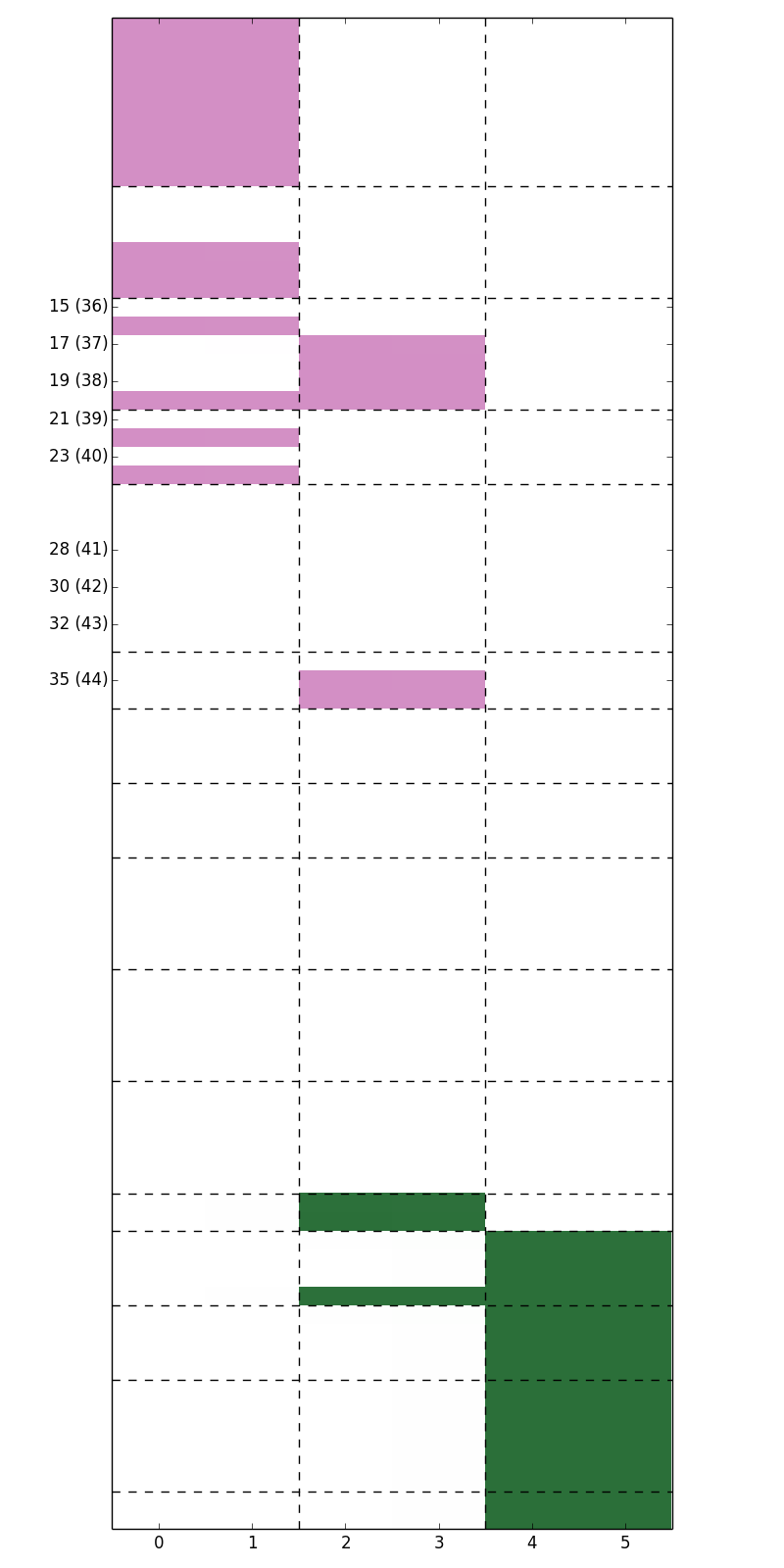}
        \caption{ghost DoFs on process 1}
        \label{fig:ghost_rows_1}
    \end{subfigure}
    \begin{subfigure}[c]{0.32\textwidth}
        \centering
        \includegraphics[width=\textwidth]{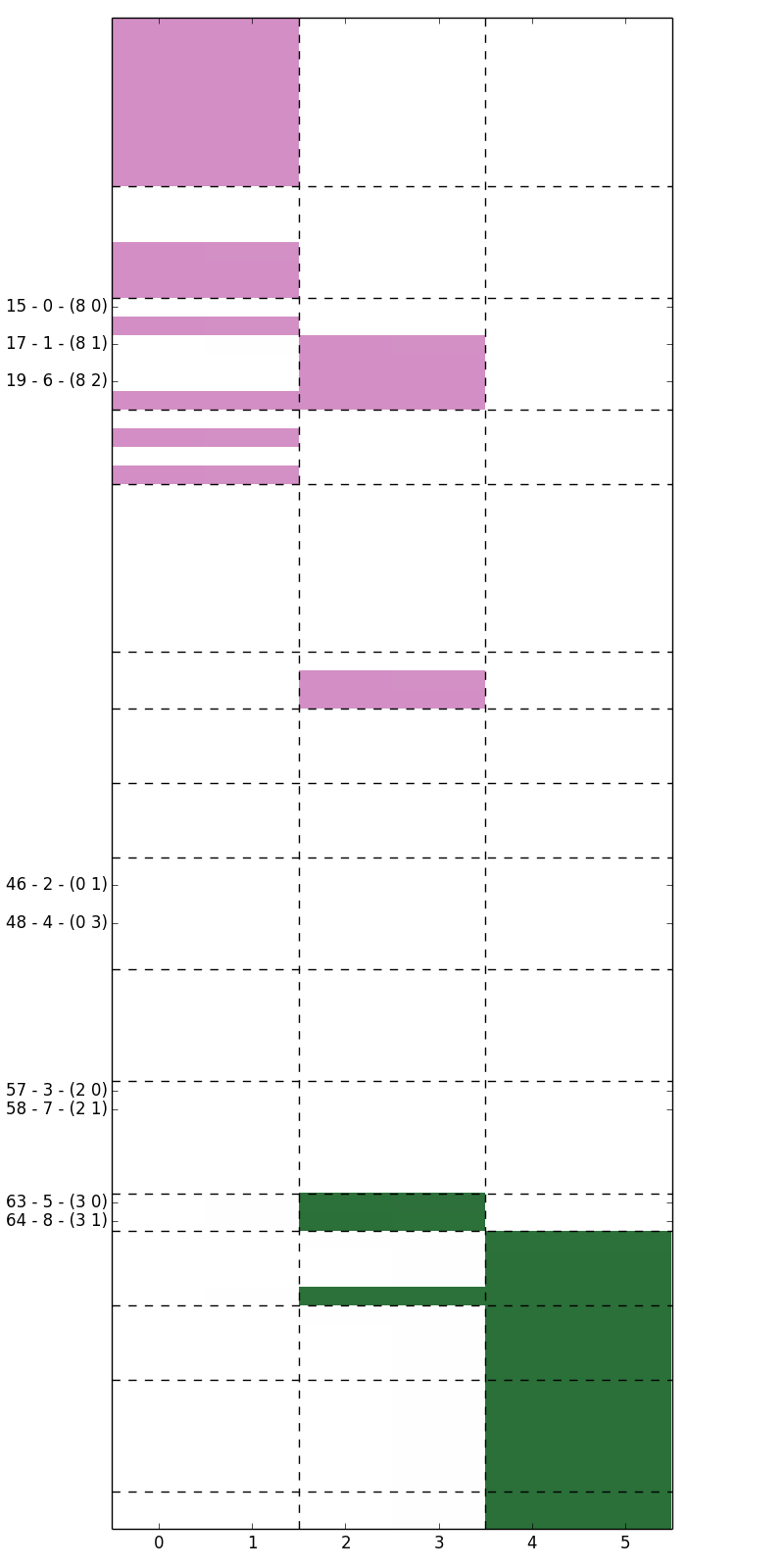}
        \caption{DoFs of a single cell in BCSR}
        \label{fig:cell_rows_1}
    \end{subfigure}
    \begin{subfigure}[c]{0.32\textwidth}
        \centering
        \includegraphics[width=\textwidth]{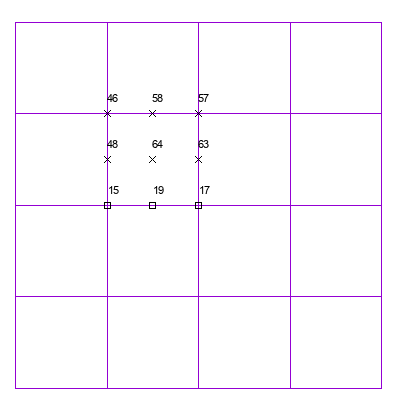}
        \caption{DoFs of a single cell}
        \label{fig:owned_proc1_cell}
    \end{subfigure}
    \caption{BCSR sparsity for the six sparse vectors with support depicted in Figure \ref{fig:fe_vectors_sparse}.
    (a) Row labels depict the global (number without brackets) and local (number in brackets) DoF numbers for ghost rows on process 1, which owns the global rows $[45,80]$ (see Figure \ref{fig:owned_proc1}).
    Following Algorithm \ref{alg:bcsr_ghosts} the depicted
    ghost rows are blocked in groups of 3 (15,17,19), 2 (21,23), 3 (28,30,32) and 1 (35).
    (b) DoFs of a single cell in BCSR structure.
    Row labels depict (i) global row number; (ii) DoF number within the cell; and (iii) block number and index within the block.
    (c) DoFs on a single cell owned by process one (see Figure \ref{fig:owned_proc1}).
    }
\end{figure}

\subsection{Memory allocation and MPI communication}

\begin{algorithm}[h]
    \SetKwInOut{Input}{Given}
    \SetKwInOut{Output}{Return}
    \Input{locally owned and ghost rows (as well as owned rows which need to be communicated with other processes upon ghost exchange, termed ``import dofs''), blocking for locally owned rows and columns, block sparsity pattern for owned block rows}
    \Output{blocking of ghost rows and their sparsity pattern}
    group ``import dofs'' according to row blocking on this MPI process \;
    collect the sparsity for such groups from locally owned block sparsity \;
    count the number of nonzero entries that will be imported for each group \;
    do point-to-point MPI communication with information about ghost blocking and their sparsity \;
    add row blocks sizes for ghost \;
    extend block sparsity for ghost row blocks \;
    \caption{Ghost blocking and sparsity pattern for multivectors.}
    \label{alg:bcsr_ghosts}
\end{algorithm}

In this section we discuss the implementation of the BCSR matrix \texttt{BlockCSRMatrix} in terms of the specifics of the \texttt{deal.II} library.
We emphasize that the ingredients are common building blocks of parallel finite element algorithms and could thus easily adopted in a different context as well.
Non-empty dense blocks in the BCSR matrix are indexed using standard CSR indexing.
In our implementation we separate the storage of matrix elements from its sparsity pattern.
The sparsity pattern is implemented as a derived class from \texttt{deal.II}'s \texttt{SparsityPatternBase}.
For each block row non-empty column block indices are stored in ascending order, which is crucial for the algorithms discussed below.
Row and column blocks are stored using the \texttt{BlockIndices} class of \texttt{deal.II}, which has been equipped with a binary search using \texttt{std::upper\_bound} when translating a global index to a block index and an index within the block.
The memory to store the non-empty blocks of sparse FE multivectors is allocated as a single 64-bytes aligned array.
Different from CSR, BCSR matrices require an additional data structure to map from the linear CSR index of the block sparsity to the beginning of the data array for this block, which we store using \texttt{std::vector}.
Similar to the design of parallel vectors, briefly introduced in Section \ref{sec:matrix_free}, \texttt{BlockCSRMatrix} adopts an ascending partitioning of the DoFs/rows with respect to the MPI rank of each process.
Local to each MPI process, ``ghost rows'' are arranged according to the rank of the MPI owner process (see Figure \ref{fig:ghost_rows_1}).
The ``update ghost values'' and ``compress'' operations are implemented similar to dense vectors using non-blocking MPI communication.
The key difference is the blocking of the ghost exchange within the matrix-free loops, i.e., the indices marked by squares on Figure \ref{fig:owned_proc1}).
Given that ghost rows are only relevant to the access pattern of current MPI process without influencing the global enumeration of the unknowns, we can block them arbitrarily.
We implement a blocking by grouping the ghost rows within the block rows of the owning MPI process, see Figure \ref{fig:ghost_rows_1} for an example.
Algorithm \ref{alg:bcsr_ghosts} lists the key ingredients to achieve this numbering.
Note that a similar strategy is applicable to the MPI partitioning of projected matrices $\overline{\gz M}$ and $\overline{\gz H}$.
In this case, however, in order to execute a local sparse matrix-matrix product we need to know not about single rows owned by other MPI processes, but about the complete row blocks.

\subsection{Matrix-free operators with sparse multivectors}
\label{sec:mf_operators}

The evaluation of discretized mass and Hamiltonian operators is implemented by a matrix-free strategy.
In this section we discuss data structures and algorithms to extend the algorithm described in Sec.~\ref{sec:matrix_free} to sparse FE multivectors stored in BCSR format.
Consider an element with DoFs as depicted in Figure \ref{fig:owned_proc1_cell}.
To identify the local degrees of freedom, we traverse the block-sparsity pattern for a given set of rows, see Figure \ref{fig:cell_rows_1}.
The local operator $\gz H_K$ must be evaluated for each column block that has a non-zero value for at least one DoF on this cell.
The identification of the relevant columns is done as follows.
We pre-compute the association from cell DoFs to row blocks and indices within the block and group this according to the block index.
This information is stored in three vectors.
The first \texttt{std::vector<std::pair<uint, uint>> dof\_indices} stores the row index within the block and the index of this DoF within the cell for all locally owned cells,
grouped by row blocks.
With the reference to Figure \ref{fig:cell_rows_1} this vector will contain
$\dots \{0,0\},\{1,1\},\{2,6\},\{1,2\},\{3,4\},\{0,3\},\{1,7\},\{0,5\},\{1,8\} \dots$.
The second \texttt{std::vector<std::pair<uint, uint>> block\_indices} stores the row block number and the number of elements within this
block for all cells, stored in an order consistent with the first one.
With the reference to Figure \ref{fig:cell_rows_1} this vector will contain $\dots \{8,3\},\{0,2\},\{2,2\},\{3,2\}\dots$.
Finally, we need a third vector
 \texttt{std::vector<std::pair<uint, uint>>}
which stores the CSR-like indices that link each cell $K$ to the beginning of the data in the other two vectors.

\begin{lstlisting}[caption={\texttt{RowBlock} class},label={lst:rowblock},language=C++]
struct RowBlock
{
    Iterator<NumberType, Constness> it;
    Iterator<NumberType, Constness> end;
    vectorized_pointer pointer;
    bool active;
    unsigned int block_row;
    ArrayView<const std::pair<uint, uint>> dof_view;
};
\end{lstlisting}

\begin{algorithm}[h]
    \SetKwInOut{Input}{Given}
    \SetKwInOut{Output}{Return}
    \Input{cell number $K$, \texttt{block\_indices} and \texttt{dof\_indices} data structures}
    \Output{setup \texttt{std::vector<RowBlock>} (see Listing \ref{lst:rowblock}) which represents active row blocks on a cell $K$ and the current column block $C$}
    \ForEach{\rm{RowBlock} on the cell $K$}{
        get iterator to the beginning of this block row from BCSR multivector \;
        get iterator to the end of this block row from BCSR multivector \;
        setup a view to \texttt{dof\_indices} for this block \;
    }
    calculate the minimum column among row blocks whose iterator does not point to ``end'' and store it in $C$ \;
    \ForEach{\rm{RowBlock}}{
        set active flag to ``true'' if its block column is $C$, and ``false'' otherwise \;
    }
    \caption{\texttt{RowsBlockAccessor::reinit()} which initializes BCSR accessor on a cell $K$}
    \label{alg:accessor_reinit}
\end{algorithm}

\begin{algorithm}[h]
    \SetKwInOut{Input}{Given}
    \SetKwInOut{Output}{Return}
    \Input{current column block number $C$, \texttt{std::vector<RowBlock>} (see Listing \ref{lst:rowblock}) which represents active row blocks on a cell $K$ for the current column block $C$}
    \Output{advanced iterators in \texttt{std::vector<RowBlock>} that are consistent with the next non-empty column block, returned by this function}
    \ForEach{\rm{RowBlock}}{
        increment the iterator if it does not point to ``end'' and its column is less than $C+1$ \;
    }
    get the minimum column among block iterators that are not at the ``end'' and store it in $C$ \;
    \If{\rm{all iterators point to ``end''}}{
        return ``invalid column number'' \;
    }
    \ForEach{\rm{RowBlock}}{
        set active flag to ``true'' if its block column is $C$, and ``false'' otherwise \;
    }
    return $C$ \;
    \caption{\texttt{RowsBlockAccessor::advance()} which iterates over non-empty column blocks on a cell $K$}
    \label{alg:accessor_advance}
\end{algorithm}

This cache-friendly data structure is used to implement the cell level read/write access to BCSR vectors in an auxiliary \texttt{RowsBlockAccessor} class.
There are two main operations: initialization of the read/write access on a cell $K$ (see Algorithm \ref{alg:accessor_reinit}) and advancement of iterators to the next non-empty column block (see Algorithm \ref{alg:accessor_advance}).
Given the row-block numbers on a cell $K$, we store iterators to the beginning of each block row in BCSR multivector (step 2 in Algorithm \ref{alg:accessor_reinit}).
Clearly not all iterators will point to the same column block (e.g. beginning iterator for row block $3$ in Figure \ref{fig:cell_rows_1} will point to the second column block) and some may even point to the end iterator for this row (e.g. row blocks $0$ and $2$ in Figure\ref{fig:cell_rows_1} are empty).
Next we determine the lowest column block number among chosen row blocks and take it as the currently active block-column $C$ (step 6 in Algorithm \ref{alg:accessor_reinit}).
We then equip each row-block iterator with a flag that is `true' if its column block is the same as currently active block-column
and `false' otherwise (step 8 in Algorithm \ref{alg:accessor_reinit}).

Given that we store the sparsity of BCSR with strictly increasing block-column indices, the iteration through non-empty column blocks for a given set of block rows is implemented straightforwardly by advancing iterators until all iterators point to the ``end'' of this block row,
see Algorithm \ref{alg:accessor_advance}.
Note that incrementing such iterators is a relatively cheap operation: a pointer to the beginning of each dense block is moved using an auxiliary vector which stores an offset for linear CSR index.

The matrix-free framework in \texttt{deal.II} has originally been developed for single vectors (i.e., matrix-vector products) \cite{Kronbichler2012}, utilizing SIMD vectorization by evaluating the operator on several cells at once via wrapper classes around intrinsics. While providing the best performance for low and moderate polynomial degrees $p\leq 10$ in the single-vector scenario \cite{Kronbichler2019}, this setup is clearly not optimal in the context of sparse FE multivectors:
when considering a batch of cells, the number of active row blocks will often be larger than on an individual cell, which implies unnecessary computations.
A natural choice for the current application is to instead perform SIMD vectorization over columns within each column block.
To achieve this goal, a row-major storage of dense blocks is desired.
Furthermore, in order to perform aligned SIMD read/write we further pad column blocks to the size of a cache line (i.e., 64 bytes).
This means that each row within each dense block is also aligned to cache boundaries.
The rationale behind this choice is to minimize the cache line transfer when reading and writing data.
As a result, we can read/write within each dense block using SIMD aligned intrinsic instructions.
An additional benefit is that we avoid a special treatment of loop remainders when reading/writing data for each row using SIMD intrinsics.
The user interface to each active row block can be conveniently combined with C++11 lambda functions
and provided for the currently active block column $C$ by \texttt{RowsBlockAccessor} class via
\begin{lstlisting}
RowsBlockAccessor::process_active_rows_vectorized([&](
  const ArrayView<const std::pair<uint,uint>> &dof_view,
  vectorized_pointer val,
  const uint stride) {
    //...
});
\end{lstlisting}
where \texttt{vectorized\_pointer} is a pointer to the beginning of the dense block in terms of \texttt{deal.II}'s wrapper class around SIMD intrinsics, called
\texttt{VectorizedArray<Number>},
\texttt{dof\_view} is a view-like object (similar to \texttt{std::span})
to the \texttt{dof\_indices} data structure described above that stores rows within the block and their index within the current cell;
and \texttt{stride} is a column stride counted in SIMD-vectorized numbers.
That is, the beginning of a row within the block can be obtained via
\texttt{val[dof\_view[i].first * stride]}.
Algorithm \ref{alg:bcsr_mf} gives a complete overview of the matrix-free operator evaluation for sparse multivectors.
Note that while applying the operator on each element, the row numbering and consequently
rows within dense blocks do not have a particular pattern. Nonetheless, no software prefetching is used as the hardware is capable of hiding access latencies.

\begin{algorithm}[h]
    \SetKwInOut{Input}{Given}
    \SetKwInOut{Output}{Return}
    \Input{sparse FE multivector $\gz U$}
    \Output{sparse FE multivector $\gz V = \gz H \gz U$}
    zero destination vector $\gz V = 0$ \;
    update ghost values for source vector $\gz U$ \;
    \ForEach{element $K \in \Omega^h$ }{
        initialize read accessor \texttt{RowsBlockAccessor} for $\gz U$ on this cell \;
        initialize write accessor \texttt{RowsBlockAccessor} for $\gz V$ on this cell \;
        \While{block column $C$ of read accessor for $\gz U$ is valid} {
            advance write accessor to point to the block column $C$ \;
            \ForEach{SIMD block in the column $C$ of a given size}{
                read non-zero DoF values of $\gz U$ using \texttt{RowsBlockAccessor} \;
                evaluate local integrals on element $K$ using sum factorization \;
                add resulting DoF values to $\gz V$ using \texttt{RowsBlockAccessor} \;
            }
            advance read accessor to the next non-empty column block \;
        }
    }
    compress operation on $\gz V$ (see Section \ref{sec:parallel_FE_vectors} and \ref{sec:parallel_sparse_FE_vectors}) \;
    \caption{Matrix-free operator with sparse multivectors stored in BCSR format.}
    \label{alg:bcsr_mf}
\end{algorithm}

The evaluation of the FE operator on a sparse multivector will produce a result vector with more non-zero elements than there are in the source vector.
This is the direct consequence of a fact that the element-wise finite element operator couples all DoF on a cell.
Given \texttt{a-priori} knowledge about the support of the source vector, it is relatively straight forward to
deduce the support for the destination vector, especially in the absence of hanging nodes constraints.

\subsection{Sparse matrix-matrix multiplication}

For completeness, this section presents the algorithms for ``mmult'' and ``Tmmult'' multiplications between BCSR matrices.
The rationale is to document straightforward implementations
including their limitations that we have adopted for the numerical examples below, rather than novel algorithms.

SpMM is a central operation for many fields, such as graph algorithms \cite{Kepner2011}, quantum mechanics calculations \cite{Borstnik2014,Rubensson2006,Challacombe2000,Rudberg2018} and algebraic multigrid preconditioners \cite{McCourt2013}.
Irregular memory access and poor data locality make SpMM a difficult kernel to optimize.
Various algorithms have been proposed in the past \cite{Buluc2008,Akbudak2018,McCourt2013, Borstnik2014,Patwary2015,Gustavson1978,Challacombe2000}.
In \cite{McCourt2013} sparse vectors were compressed into full vectors using graph coloring and used in sparse matrix dense multivector product.
A modification of the Canon algorithm for 2D MPI-partitioned matrices in doubly compressed sparse column format was proposed in \cite{Buluc2008}.
BCSR matrices have been used for quantum mechanics calculations with atom-centered basis functions in \cite{Challacombe2000,Saravanan2003,Borstnik2014}.

In the present context, multiplication between tall and skinny sparse multivectors is necessary.
The unknown solution vector $\gz \Phi$ employs a 1D row-wise MPI partitioning associated with the
FE discretization of the domain $\Omega$, as discussed above. Furthermore, the efficient implementation of the matrix-free operator evaluation with fast read/write access to each non-zero DoF on a cell
motivated our choice of BCSR matrix format
as opposed to quadtree representation of matrices \cite{Bock2016,Bock2013} or hierarchical matrices \cite{Rubensson2006,Rubensson2007,Rudberg2018}.
We note that projected matrices (denoted by $\overline{\{\bullet \}}$ in Section \ref{sec:required_op})
are relatively small compared to the unknown solution vectors $\gz \Phi$.
This suggests adopting 1D row-wise MPI partitioning for these matrices as well.

\begin{algorithm}[h]
    \SetKwInOut{Input}{Given}
    \SetKwInOut{Output}{Return}
    \Input{source BCSR matrices $\gz A$ and $\gz B$, destination matrix $\gz C$ with given sparsity pattern}
    \Output{$\gz C = \gz A \gz B$}
    zero destination matrix $\gz C = 0$ \;
    update ghost values for matrix $\gz B$ \;
    \ForEach{locally owned row block $i$}{
        set $A_i$ to the iterator to the beginning of the row block $i$ in $\gz A$ \;
        \While{$A_i$ does not point to the ``end'' of this row block} {
            set $k$ to the column block of $A_i$ \;
            set $B_k$ to the iterator to the beginning of the row block $k$ \;
            reset $C_i$ to the iterator to the beginning of the row block $i$ in $\gz C$ \;
            \While{$B_k$ does not point to the ``end'' of this row} {
                set $j$ to the block column of $B_k$ \;
                advance $C_i$ iterator until its column is more or equal than $j$ \;
                \If{column of $C_i$ is equal to $j$}{
                    dense matrix-matrix multiplication for blocks via \texttt{dgemm} \;
                }
                advance $B_k$\;
            }
            advance $A_i$ \;
        }
    }
    \caption{Matrix-matrix multiplication for BCSR format using three nested loops. The inner loop allows for truncation in case sparsity of the destination matrix does not contain the product.}
    \label{alg:bcsr_mmult}
\end{algorithm}

A matrix-matrix product $\gz C = \gz A \gz B$ involves three nested loops.
An outer loop runs over the rows of $\gz A$, for each row we then loop over all
the columns, and then we need to multiply each element with all the
elements in that row in $\gz B$, see Algorithm \ref{alg:bcsr_mmult}.
Note that $\gz A$ and $\gz C$ are assumed to have the same row partitioning, whereas
$\gz B$ should have ghost values consistent with the sparsity of $\gz A$ on this MPI process.
That is, step 7 of Algorithm \ref{alg:bcsr_mmult} may result in access outside of locally owned row blocks of $\gz B$.
Therefore ghost values of $\gz B$ have to be communicated before the evaluation of the product.
Additionally we also allow for an inexact product, that is, when the sparsity of $\gz C$ is not enough to store the product and some entries are discarded.
This is a typical scenario for OM presented in Section \ref{sec:required_op} as gradients of the functional $\gz G$ are more dense
than the unknown multivectors $\gz \Phi$.

\begin{algorithm}[h]
    \SetKwInOut{Input}{Given}
    \SetKwInOut{Output}{Return}
    \Input{source BCSR matrices $\gz A$ and $\gz B$, destination matrix $\gz C$ correct sparsity pattern}
    \Output{$\gz C = \gz A^T \gz B$}
    zero destination matrix $\gz C = 0$ \;
    \ForEach{locally owned row block $i$ in $\gz A$}{
        set $A_i$ to the iterator to the beginning of the row block $i$ in $\gz A$ \;
        \While{$A_i$ does not point to the ``end'' of this row block} {
            set $k$ to the column block of $A_i$ \;
            set $C_k$ to the iterator to the beginning of the row block $k$ in $\gz C$ \;
            reset $B_i$ to the iterator to the beginning of the row block $i$ in $\gz B$ \;
            \While{$B_i$ does not point to the ``end'' of this row} {
                set $l$ to the block column of $B_i$ \;
                advance $C_k$ iterator until its column equals $l$ \;
                dense matrix-matrix multiplication for blocks via \texttt{dgemm} \;
                advance $B_i$\;
            }
            advance $A_i$ \;
        }
    }
    ``compress'' operation on $\gz C$ (see Section \ref{sec:parallel_FE_vectors} and \ref{sec:parallel_sparse_FE_vectors}) \;
    \caption{Transpose matrix-matrix multiplication for BCSR format using three nested loops.}
    \label{alg:bcsr_Tmmult}
\end{algorithm}

Algorithm \ref{alg:bcsr_Tmmult} presents our implementation of the transpose matrix-matrix multiplication.
The product involves three nested loops in analogy to the non-transposed multiplication. Note
that block rows of $\gz C$ are accessed in a non-uniform way, which requires synchronization at the end of the local products, a step we call (``compress'').

\subsection{A summary of new functionality added to deal.II}
In this section we present a list of classes and methods that are being added to the open source library \texttt{deal.II}
\begin{itemize}
\item \texttt{Utilities::inverse\_Hilbert\_space\_filling\_curve()} assigns to each point \texttt{Point<dim,double>} in \texttt{dim} spacial dimensions an index  \texttt{std::array<std::uint64\_t,dim>} using the Hilbert space filling curve;
\item \texttt{BlockCSRMatrix} that implements \texttt{BCSR} matrix, as well as matrix-matrix products and MPI communication for ``compress'' and ``update ghosts'' operations;
\item \texttt{RowsBlockAccessor} an auxiliary class that provides read/write access to rows in \texttt{BlockCSRMatrix} during matrix-free FE operator evaluation;
\item various improvements and extensions to \texttt{DynamicSparsityPattern} and \texttt{SparsityPattern} classes, which we rely upon in implementing \texttt{BCSR} matrix and its iterators;
\end{itemize}
The algorithms  developed here will be made publicly available as a part of the deal.II library.

\section{Numerical results}
\label{sec:numerical_results}

\begin{figure}[!ht]
    \centering
    \includegraphics[width=0.5\textwidth]{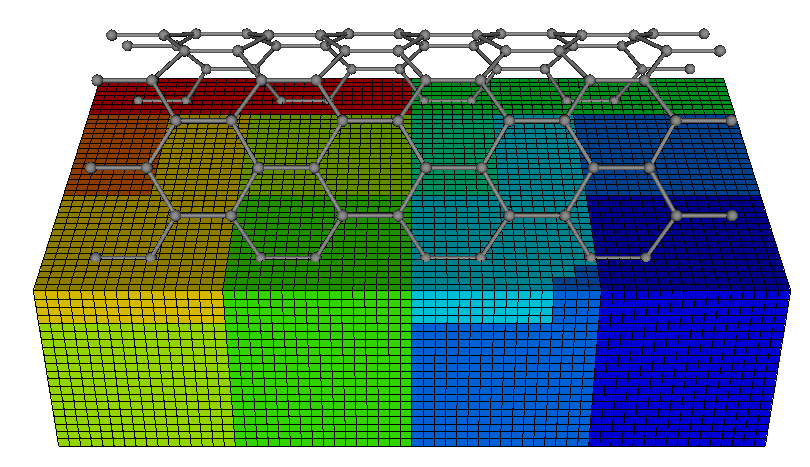}
    \caption{The cut of the mesh for $M=320$ (carbon nanotube with $160$ atoms) with quadratic FE basis.
    Color indicates MPI partitioning into ten processes.}
    \label{fig:mesh_C160}
\end{figure}

As a benchmark problem, we look at carbon nanotubes of different lengths.
Note that within this contribution we do not solve the DFT but only benchmark and study the required operations for the OM approach.
Carbon nanotubes are chosen as a benchmark since these systems are suitable for OM methods
and they are convenient for scaling studies in a strictly reproducible way.
Nanotubes with $(10,0)$ chirality were generated by the TubeGen \cite{TubGen34} tool and consist of  $160$, $320$, $640$, $1280$, $2560$, $5120$, $10240$ atoms.
We assign two sparse vectors to each Carbon atom.
Each vector has support within the distance of $8$ Bohr from the corresponding atom \cite{Fattebert2000}.
In order to have full control over the meshing, we choose a simple structured mesh
and make sure that the number of elements (and therefore the number of DoFs)
is proportional to the number of atoms.
The sparsity of multivectors in terms of cell patches is constructed on a mesh level one coarser than the finest level
based on the proximity of cell centers to each orbital.
For each patch we can construct the sparsity in terms of DoFs by taking shape functions $N_i(\gz x)$
that have support only within the patch (see Figure \ref{fig:fe_vectors_sparse} for a 2D example).
The same mesh level is used for the enumeration and blocking of the DoFs using HSFC.
The blocking parameter $B$ for orbitals in Algorithm \ref{alg:vec_renumbering} is set to $8$.
Judging from the arithmetic complexity of the matrix-free approach (see Figure \ref{fig:complexity}), one can argue that for scalar operators the method is especially attractive for degrees between two and six, where the cost per unknown is lowest.
For this benchmark we choose quadratic and quartic polynomial bases.
The smallest element size (in Bohr) for the two cases are $0.5625\times0.5\times0.5$ and $1.125\times1\times1$, respectively.
The length of the mesh for each nanotube is scaled proportionally to the number of atoms.
As an example, Figure \ref{fig:mesh_C160} shows the mesh for the $\rm C160$ nanotube used with quadratic FEs.
The crossection of the mesh is $40\times40$ square.

Unless noted otherwise, the numerical experiments are performed on a cluster with each node having two Xeon 2660v2 ``Ivy Bridge'' chips (10 cores per chip + SMT) running at 2.2 GHz (turbo boost disabled) with 25 MB Shared Cache per chip and 64 GB of RAM.
The code has been run with
\begin{itemize}
\item the Intel compiler version 18.03 with flags ``-O3 -march=native'', Intel MPI version 18.03 and Intel MKL 2018.3; and
\item the GCC compiler version 8.2.0 with flags ``-O3 -march=native'', OpenMPI version 3.1.3 and Intel MKL 2018.3.
\end{itemize}
The memory bandwidth and the peak performance for a single socket are 47.2 GBytes/s and 176 GFlop/s, respectively.

\subsection{Comparison of sparse multivectors to dense column vectors}

\begin{table}
    \centering
    \begin{tabular}{|r|c|c|}
    \hline
                    & quadratic & quartic \\
    \hline
    DoFs per cell   & \dataptwodofspercell & \datapfourdofspercell \\
    number of cells & \dataptwoncells      &  \datapfourncells \\
    DoFs (=rows)    & \dataptwodofs        & \datapfourdofs \\
    columns         & \dataptwocolumns     & \datapfourcolumns \\
    number of row blocks & \dataptwonrowblocks & \datapfournrowblocks \\
    number of column blocks & \dataptwoncolblocks & \datapfourncolblocks \\
    row blocks size & \pgfmathprintnumber{\dataptworowblocksmean} $\pm$ \pgfmathprintnumber{\dataptworowblocksstd} & \pgfmathprintnumber{\datapfourrowblocksmean} $\pm$ \pgfmathprintnumber{\datapfourrowblocksstd} \\
    column blocks size & \pgfmathprintnumber{\dataptwocolblocksmean} $\pm$ \pgfmathprintnumber{\dataptwocolblocksstd} & \pgfmathprintnumber{\datapfourcolblocksmean} $\pm$ \pgfmathprintnumber{\datapfourcolblocksstd} \\
    locally owned row blocks in $\overline{\gz H}$ & \pgfmathprintnumber{\dataptwocolblocksmpimean} $\pm$ \pgfmathprintnumber{\dataptwocolblocksmpistd} & \pgfmathprintnumber{\datapfourcolblocksmpimean} $\pm$ \pgfmathprintnumber{\datapfourcolblocksmpistd} \\
    non-empty column blocks for FE cell op. & \pgfmathprintnumber{\dataptwomfnonemtpycolumnblocksmean} $\pm$ \pgfmathprintnumber{\dataptwomfnonemtpycolumnblocksstd} & \pgfmathprintnumber{\datapfourmfnonemtpycolumnblocksmean} $\pm$ \pgfmathprintnumber{\datapfourmfnonemtpycolumnblocksstd} \\
    \hline
    \end{tabular}
    \caption{Statistics for the smallest example with $160$ atoms depicted in Figure \ref{fig:mesh_C160} run on 10 MPI processes.}
    \label{tab:bcsr_statistics}
\end{table}

First we compare our implementation in the BCSR format to a setup using full column vectors for the smallest example with $160$ atoms run on a single socket (i.e. with 10 MPI processes).
Table~\ref{tab:bcsr_statistics} reports various information for this setup.
Clearly tall and skinny sparse multivectors result in dense blocks of the same character.
Note that most of the column blocks are of size $8$ (as expected),
which indicates that the matrix-free cell operator will require two SIMD swipes per column block on average (see Step 8 in Algorithm \ref{alg:bcsr_mf}).
Additionally the table reports the number of non-empty column blocks accumulated over all cells during application of the matrix-free operator
to sparse multivector $\gz \Phi$, see Step 6 of Algorithm \ref{alg:bcsr_mf}.
The variation in this number can be considered as a measure of work imbalance for this kernel.

\begin{table}
    \centering
    \begin{tabular}{|r|rr|rr|}
    \hline
                    &       \multicolumn{2}{c|}{quadratic}              &          \multicolumn{2}{c|}{quartic}                \\
                    & BCSR                    & Full                    & BCSR                     &  Full                     \\
    \hline
    FE operator [s] & \pgfmathprintnumber{\dataptwoWallHBcsr}      & \pgfmathprintnumber{\dataptwoWallHFull}      & \pgfmathprintnumber{\datapfourWallHBcsr}      & \pgfmathprintnumber{\datapfourWallHFull}       \\
    mmult [s]       & \pgfmathprintnumber{\dataptwoWallmmultBcsr}  & \pgfmathprintnumber{\dataptwoWallmmultFull}  & \pgfmathprintnumber{\datapfourWallmmultBcsr}  & \pgfmathprintnumber{\datapfourWallmmultFull}   \\
    Tmmult [s]      & \pgfmathprintnumber{\dataptwoWallTmmultBcsr} & \pgfmathprintnumber{\dataptwoWallTmmultFull} & \pgfmathprintnumber{\datapfourWallTmmultBcsr} & \pgfmathprintnumber{\datapfourWallTmmultFull}  \\
    Memory [Mb]     & \pgfmathprintnumber{\dataptwobcsrmemory}     & \pgfmathprintnumber{\dataptwofullmemory}     & \pgfmathprintnumber{\datapfourbcsrmemory}     & \pgfmathprintnumber{\dataptwofullmemory}       \\
    \hline
    \end{tabular}
    \caption{Comparison of the BCSR-based implementation to the dense column vectors compiled with Intel compiler.}
    \label{tab:bcsr_vs_full}
\end{table}

Table \ref{tab:bcsr_vs_full} compares the run time and memory consumption of an implementation in the BCSR format versus a setup using full column vectors.
Besides a reduction in memory of over $3\times$ (measured as the resident memory reported by Linux), the run times for all components are also considerably reduced.
Clearly, this is most obvious for the matrix-matrix multiplications.
However, also the matrix-multivector product is more than 3.5 times faster, verifying the efficiency of the implementation.
For larger configurations with more atoms, the advantage of the BCSR-based implementation grows, and a setup with full column vector would soon become intractable.

\subsection{Node-level performance}

\begin{figure}[!ht]
    \begin{subfigure}[b]{0.49\textwidth}
        \centering
        \includegraphics[width=\textwidth]{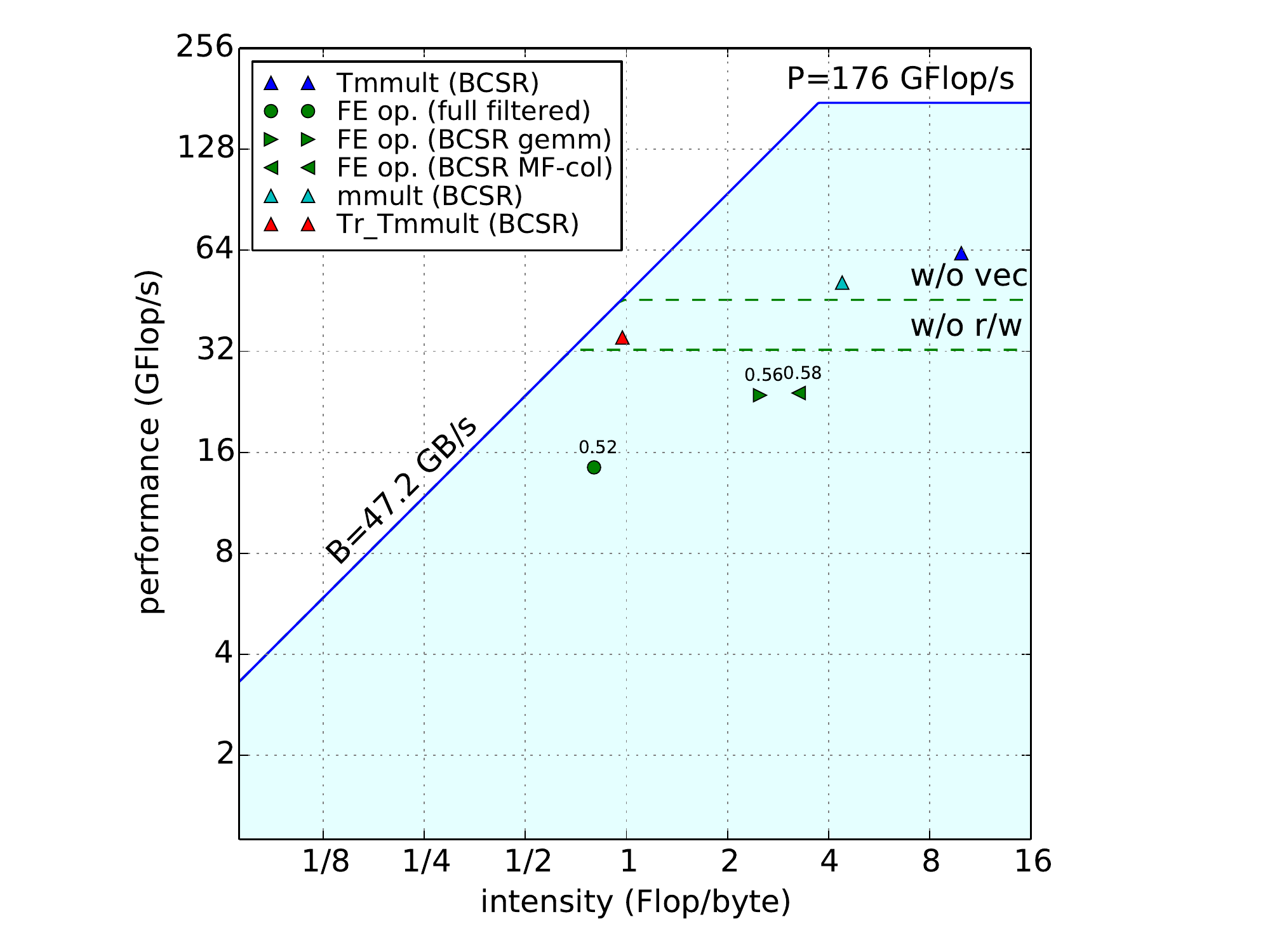}
        \caption{Quadratic elements (Intel)}
        \label{fig:roofline_2}
    \end{subfigure}
    ~
    \begin{subfigure}[b]{0.49\textwidth}
        \centering
        \includegraphics[width=\textwidth]{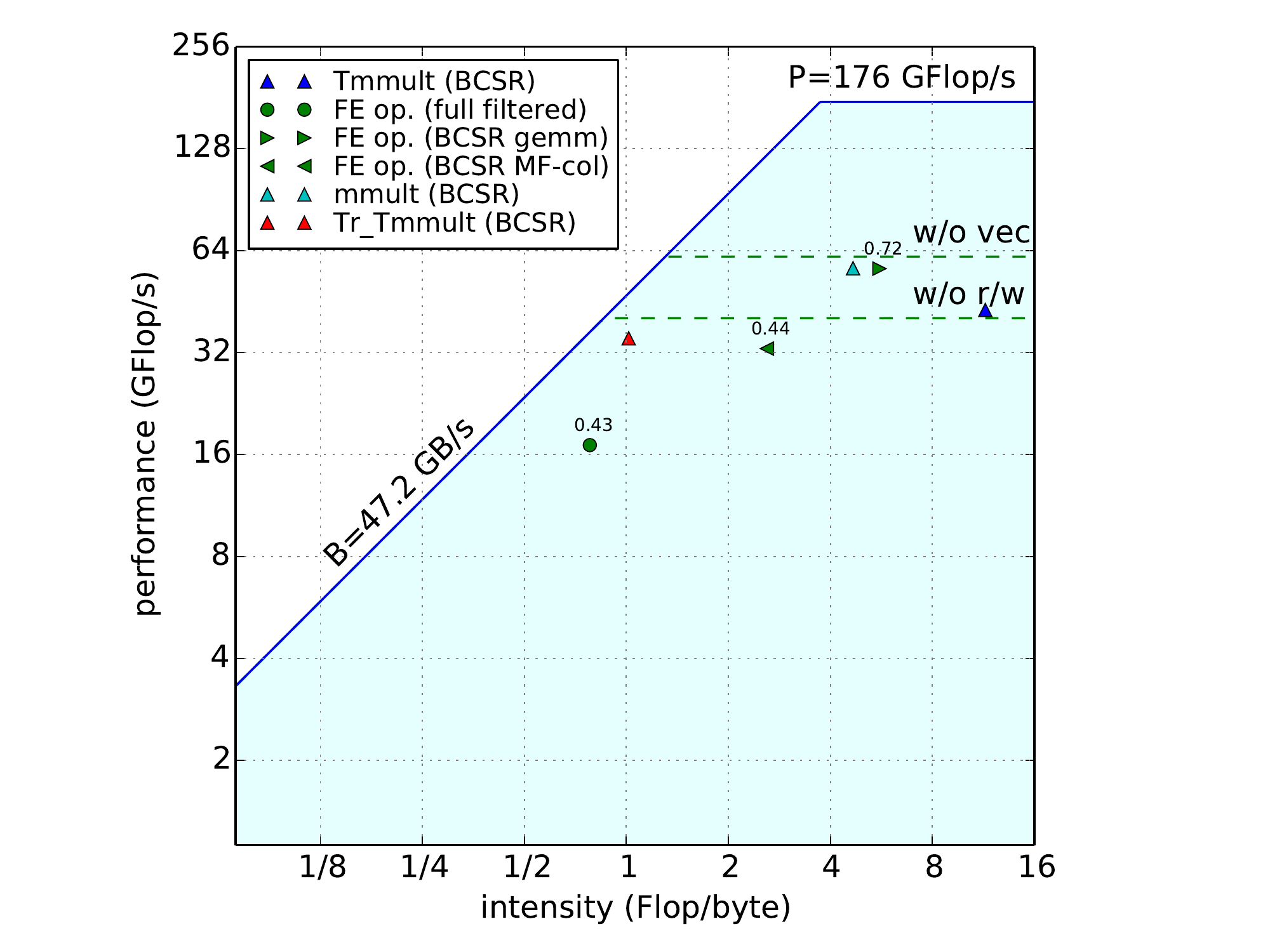}
        \caption{Quartic elements (Intel)}
        \label{fig:roofline_4}
    \end{subfigure}
    \\
    \begin{subfigure}[b]{0.49\textwidth}
        \centering
        \includegraphics[width=\textwidth]{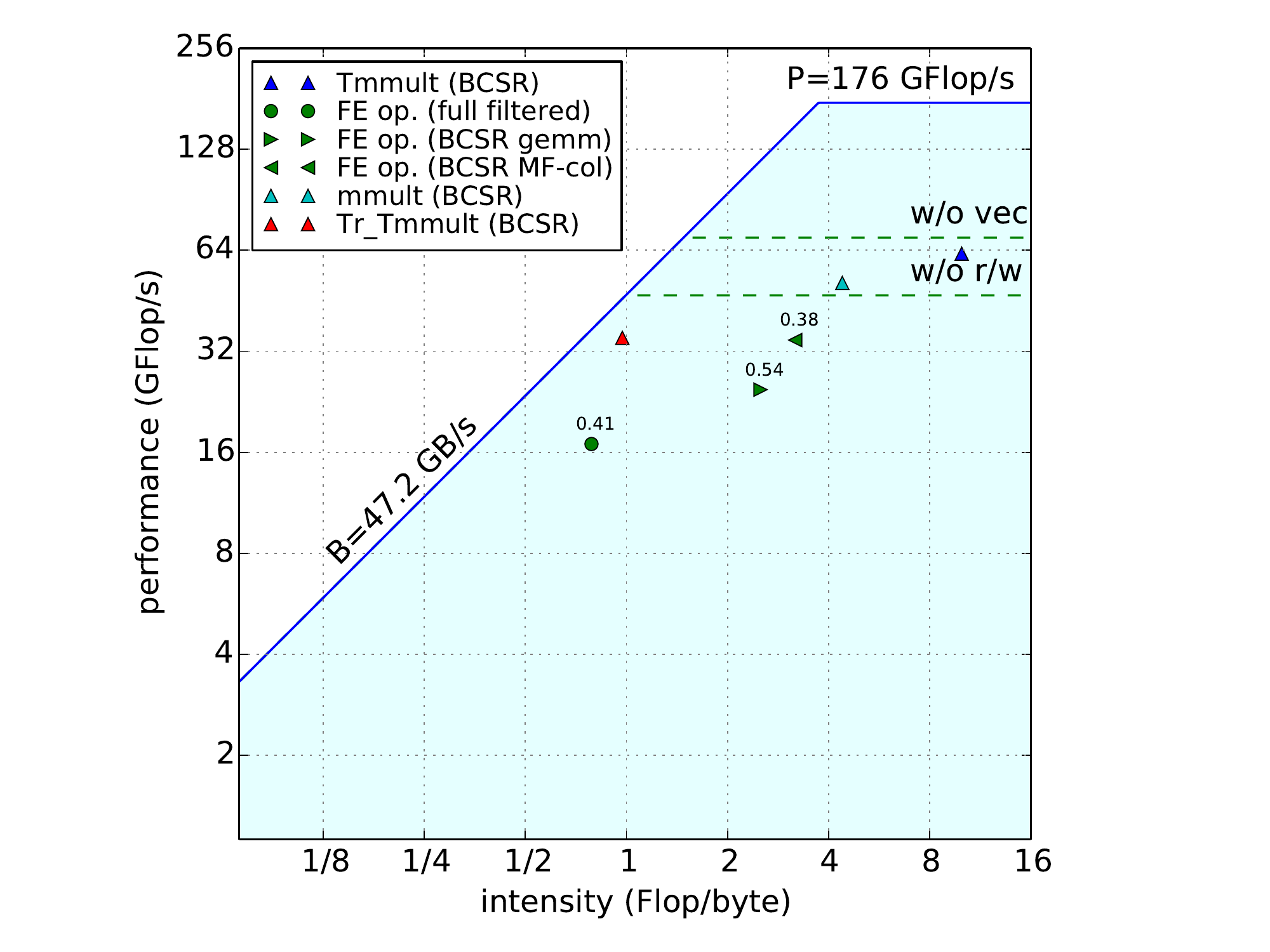}
        \caption{Quadratic elements (GCC)}
        \label{fig:roofline_2gcc}
    \end{subfigure}
    ~
    \begin{subfigure}[b]{0.49\textwidth}
        \centering
        \includegraphics[width=\textwidth]{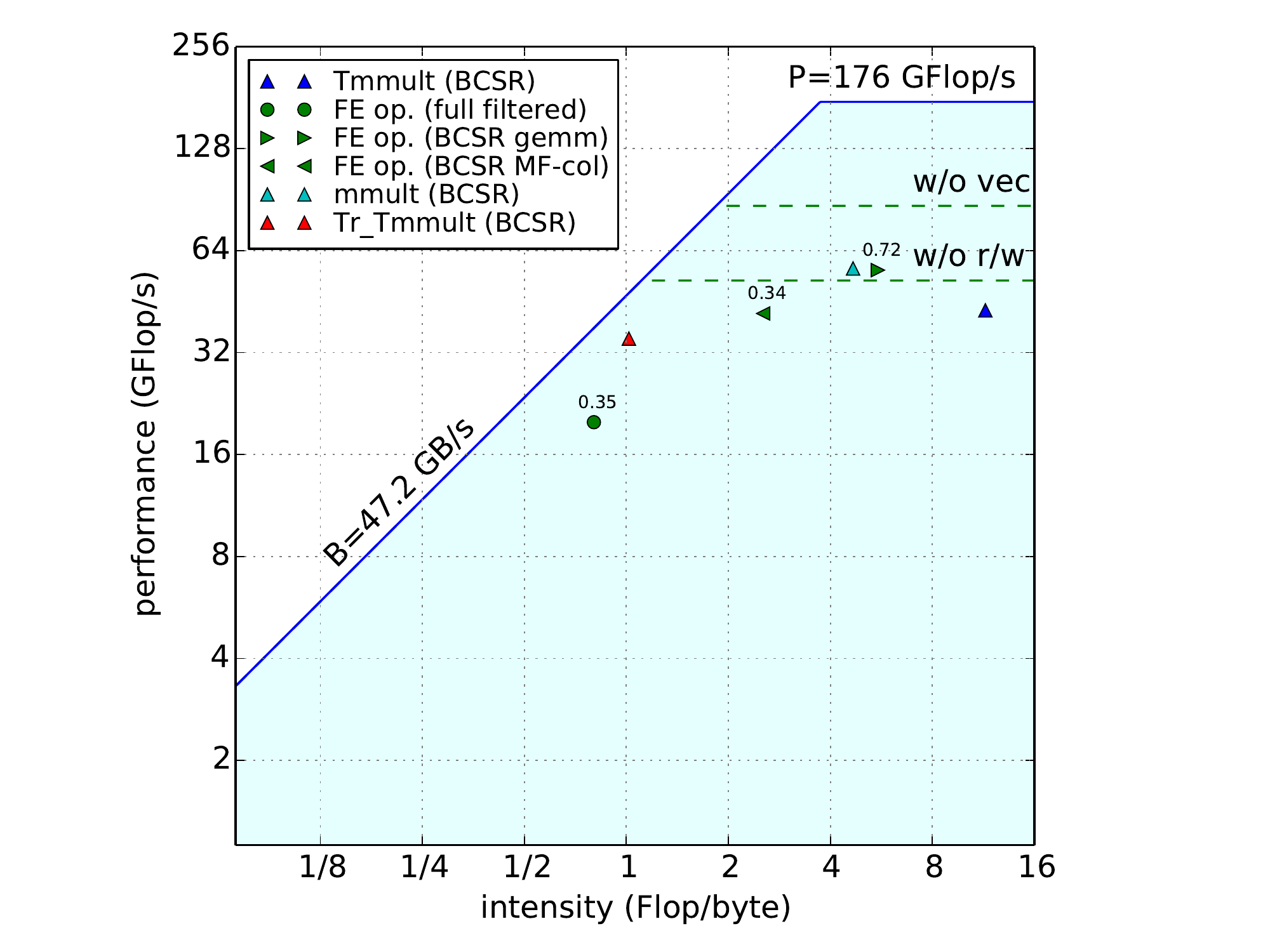}
        \caption{Quartic elements (GCC)}
        \label{fig:roofline_4gcc}
    \end{subfigure}
    \caption{Roofline performance model.
    The dashed roof line "w/o r/w" represents the matrix-free Algorithm \ref{alg:bcsr_mf} without reading from and writing into multivectors (i.e., steps 9 and 11),
    whereas "w/o vec" represents Algorithm \ref{alg:bcsr_mf} without any multivector operations (i.e., steps 1,2,9,11 and 16). The absolute run time per call for the three matrix-free flavors is given by numbers next to the respective symbols.
    }
    \label{fig:roofline}
\end{figure}

Next we analyze the node-level performance of our kernels for BCSR sparse FE vectors using the roofline model
$p = \rm{min}(P, B I)$,
where $P$ is the peak arithmetic performance, $B$ is the peak main memory bandwidth and $I$ is the computational intensity.
The sustained memory bandwidth (MBytes/s) and the number of floating point operations per second (MFLOP/s) are measured using the MEM DP group of the LIKWID \cite{Treibig2010} hardware performance counter tool, version 4.3.3.
The measurements are done on a single socket of the RRZE cluster by pinning all 10 MPI processes to this socket.
The smallest example with $160$ atoms is considered.
All results are reported for double precision numbers with explicit SIMD vectorization over 4 lanes organized according to Section~\ref{sec:mf_operators}.
Walltime is reported as an average of each kernel executed ten times in a row.

Figure \ref{fig:roofline} shows results of the roofline model for the considered kernels for quadratic and quartic FE bases.
The mesh for quadratic FEs is depicted in Figure \ref{fig:mesh_C160}, whereas the quartic mesh is coarser by one level, which results in the same number of total DoFs, see Table \ref{tab:bcsr_statistics}.
Additionally we compare the matrix-free approach to cell-level \texttt{dgemm} operation using BLAS, labeled ``FE op. (BCSR gemm)'', according to the recent open-source implementation of DFT using the \texttt{deal.II} library \cite{Motamarri2019},
as well as dense column vectors with filtering of the matrix-free FE operator based on support for each vector, labeled ``FE op. (full filtered)''.
In order to facilitate comparison we also report the wallclock time for the three flavors of the matrix-free operators in the roofline diagram.

\begin{figure}[!ht]
    \begin{subfigure}[b]{0.49\textwidth}
        \centering
        \includegraphics[width=\textwidth]{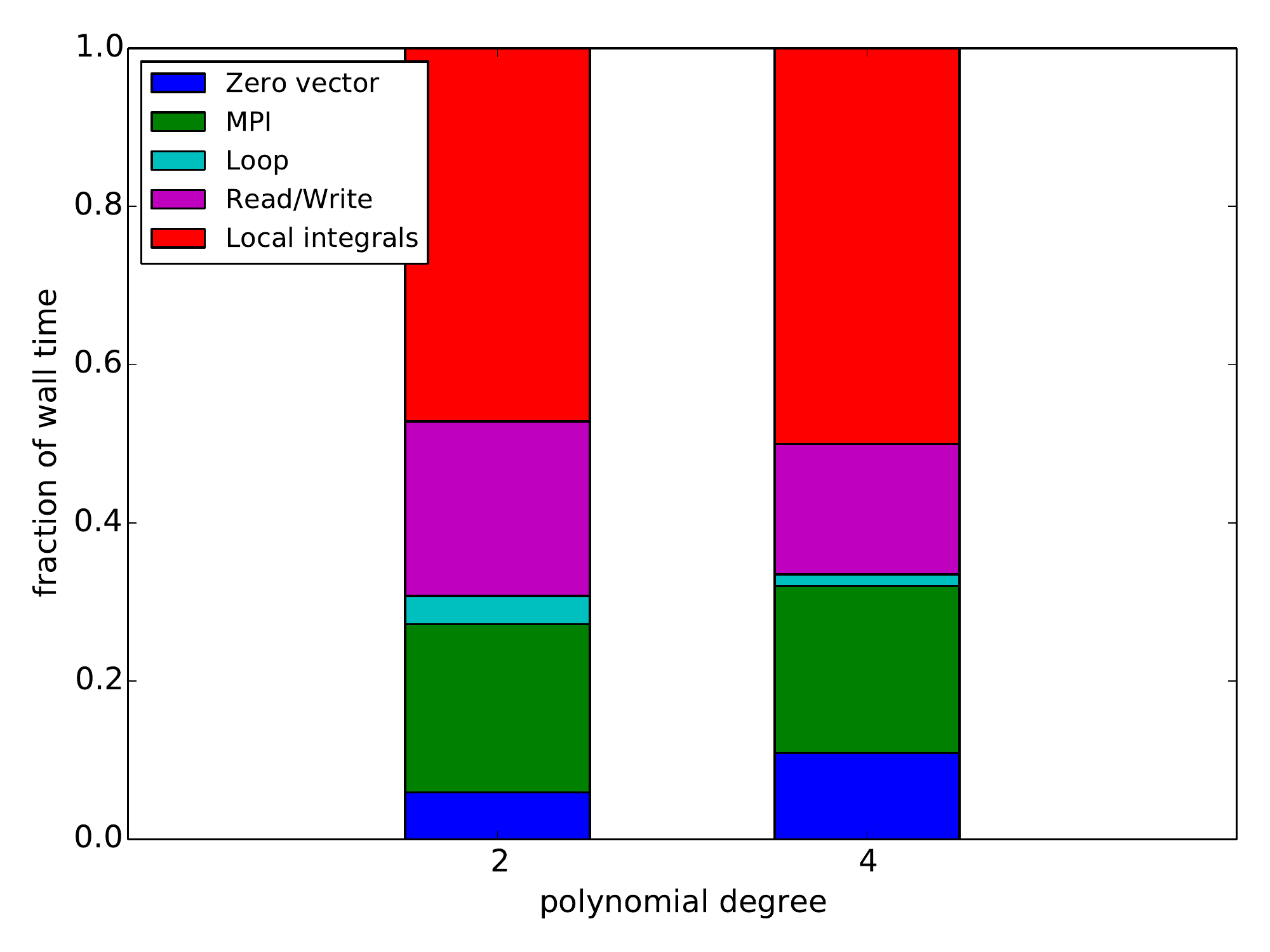}
        \caption{Intel}
        \label{fig:breakdownIntel}
    \end{subfigure}
    ~
    \begin{subfigure}[b]{0.49\textwidth}
        \centering
        \includegraphics[width=\textwidth]{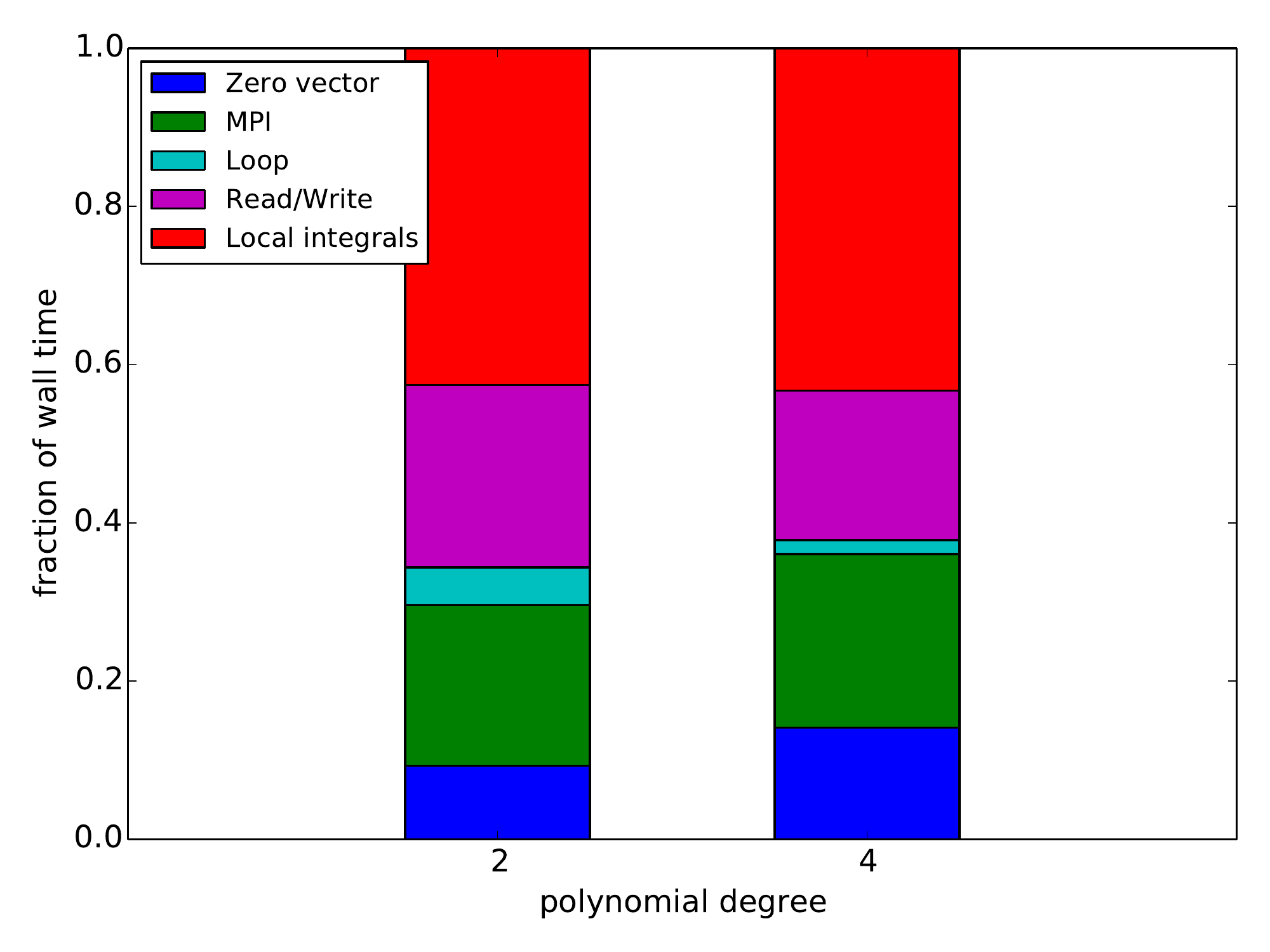}
        \caption{GCC}
        \label{fig:breakdownGCC}
    \end{subfigure}
    \caption{Breakdown of computing times of various steps of Algorithm \ref{alg:bcsr_mf} on 10 cores.}%
    \label{fig:breakdown}
\end{figure}

The sparse matrix-matrix products (``mmult'', ``Tmmult'' and ``Tr\_tmmult'') are in the core bound regime.
By profiling these kernels we observe that more than 90\% of time is spent in the BLAS \texttt{dgemm} function, showing that no significant overhead is introduced by our implementation.
For the matrix-vector products, the matrix-free operator evaluation is considerably faster than the cell-wise full matrices for the quartic basis.
This is expected as there is a large reduction in arithmetic complexity for higher order elements by sum factorization, see Figure \ref{fig:complexity}.
However, both operators are relatively far away from the arithmetic peak performance.
We also observe that a basic matrix-free evaluation with filtered column vectors, labeled ``FE op. (full filtered)'', involves a lower arithmetic throughput as well as a lower computational intensity,
albeit for a comparable run time to the BCSR matrix-free implementation. For this setup
the memory for dense column vectors is allocated regardless of the underlying sparsity and thus the
indirections in the index access are avoided for ``FE op (full filtered)''. Apart from the filter and some associated additional memory traffic (which can be related to overhead in the index storage as well as eager prefetching of filtered-out vector data), this data point represents the best case for the single-vector matrix-free operation evaluation provided by the \texttt{deal.II} library.
The increased arithmetic intensity confirms the beneficial properties of the column-oriented algorithm in the BCSR format in general and our implementation in particular for matrix-free computations on sparse multivectors.

When comparing results obtained by using different compilers, we observe that
the GCC compiler gives  30-40\% higher performance for the matrix-free FE operator (%
\pgfmathprintnumber{\dataGCCptwoGFlopsHBcsr}
vs
\pgfmathprintnumber{\dataptwoGFlopsHBcsr}
GFLOP/s for quadratic FEs and
\pgfmathprintnumber{\dataGCCpfourGFlopsHBcsr}
vs
\pgfmathprintnumber{\datapfourGFlopsHBcsr}
for quartic FEs
).
This aligns with our previous observations that the GCC compiler generates better machine code for
the sum factorization algorithms implemented within \texttt{deal.II}
\cite[Sec. 3]{Kronbichler2019}.
Note that Intel's MKL is used for the BLAS and LAPACK interface with both compilers, so the performance of the matrix-matrix products, as expected, is independent of the utilized compiler.
GCC compiler leads to a higher a memory throughput
(e.g., \pgfmathprintnumber{\dataGCCptwoBandWidthHBcsr} GB/s vs
\pgfmathprintnumber{\dataptwoBandWidthHBcsr} GB/s for quadratic FEs).

Details about the various stages of the BCSR matrix-free algorithm in ``FE op. (BCSR MF-col)'' of Figure~\ref{fig:roofline} are listed in Figure \ref{fig:breakdown} by a stack plot.
Since adding timers inside the element loop (step 3 of the algorithm) can severely change the total runtime due to timer overheads, we measured the time indirectly as follows.
Using a template parameter in C++, we selectively disable various stages of the algorithm.
The following variations of
Algorithm \ref{alg:bcsr_mf} are considered:
(i) element loop without read/write operations (steps 9 and 11);
(ii) element loop without evaluating local integrals (step 10);
(iii) element loop (step 3);
(iv) the complete algorithm without resetting destination vector to zero (step 1);
(v) the complete algorithm; and
(vi) the element loop without read/write operations and local integrals (step 9, 10 and 11)\footnote{We prevented the compiler from
optimizing away the loop at step 8 using inline assembly language.
}.

From these measurements we can reliably deduce the wall-clock time for five key steps plotted in Figure \ref{fig:breakdown}:
(i) resetting destination vector (step 1);
(ii) MPI communication (steps 2 and 16);
(iii) read from and write into BCSR vectors (steps 9 and 11);
(iv) evaluation of local integrals (step 10);
(v) cell loop and BCSR accessors overhead (steps 3-8, 13).
The measurements show that for quadratic elements and the code compiled with Intel compiler
\dataptwoStackIntegrals\% of the time is spent in the local evaluation of integrals, i.e., the product $\gz H_K \gz U_K$ done via quadrature with sum factorization, whereas \dataptwoStackReadWrite\% is spent reading from/writing to BCSR vectors.
\dataptwoStackMPI\% of the time is spent in MPI communication.
Note that this step also includes reading from and writing to auxiliary
vectors in terms of pack/unpack operations within the non-blocking MPI
communication. While this proportion might seem high, it is a consequence
of the code optimizations in the other parts, in particular for the local
integrals and the BCSR data access, see also \cite{Kronbichler2019} where
a similar observation was made. We also emphasize that overlapping of
communication and computation is not applicable here as the slowdown is
in fact observed within the shared-memory region of a single socket where
both the pack/unpack as well as the actual exchange between MPI contribute
with similar shares.
Most importantly, the loop overhead for BCSR multivectors only takes \dataptwoStackLoop\% of the walltime,
which indicates that our implementation of row iterators for BCSR in \texttt{RowsBlockAccessor} adds little overhead to the overall algorithm.
Results for quartic elements are largely similar.
When comparing the measurements for quadratic FE obtained with the Intel compiler to those
obtained by employing the GCC compiler (see Figure \ref{fig:breakdownGCC}), we observe that the
local evaluation of integrals takes a smaller fraction of time (\dataGCCptwoStackIntegrals\%).
Note that the BCSR loop overhead for this case is still remarkably low -- only \dataGCCptwoStackLoop\% of the walltime. Finally, we emphasize that the share of the ``Loop'' and ``MPI'' parts in Figure~\ref{fig:breakdown} is also affected by the slight load imbalance as some MPI ranks must wait for neighbors that still perform local computations in the ghost exchange.

In order to identify the influence of the sparse vectors on the sub-optimal performance, we plot two additional data points in the roofline diagram, plotted as dashed horizontal lines because the arithmetic intensity is partly outside of the picture.
The lower line reports the performance of the matrix-free operator without reading from and writing into the sparse vectors (steps 9 and 11 in Algorithm \ref{alg:bcsr_mf}).
The measured actual matrix-free kernel is relatively close to this roof of \pgfmathprintnumber{\dataptwoGFlopsHBcsrNoRW} GFLOP/s (Intel compiler).
In other words, data movement for BCSR multivectors is shown to be efficient.
The second line is obtained by additionally disabling the zero-out of the source vector and the MPI communication (steps 1,2 and 16 in Algorithm \ref{alg:bcsr_mf}).
Thus, this line represents the local cell operations for BCSR vectors without any data access except the quadrature point data.
For quadratic elements its performance value is
\pgfmathprintnumber{\dataptwoGFlopsHBcsrMFOnly} GFLOP/s with the Intel compiler, corresponding to one fourth of the peak performance.
The minimum and maximum walltime for this stage of the algorithm are
\pgfmathprintnumber{\dataptwoWallMinHBcsrMFOnly}s
and
\pgfmathprintnumber{\dataptwoWallHBcsrMFOnly}s,
respectively.
This work imbalance is caused by the variation in the total
number of non-empty column blocks for all cells during application of the matrix-free operator, see the last row of Table \ref{tab:bcsr_statistics}.
Therefore, future studies with a focus on better balancing between the cost of the MPI data exchange, the local integrals as well as the read/write operations could bring the overall algorithm somewhat closer to this arithmetic.

Matrix-free FE operators with BCSR multivectors potentially perform a different amount of work on each cell. Consequently it is not possible to fuse work over multiple FE cells.
As an alternative we tried to pipeline work within the inner most loop in the matrix-free kernel (step 8 in Algorithm \ref{alg:bcsr_mf}) for column blocks of sizes larger than or equal to 8 (SIMD vectorization is over 4 doubles).
In particular one can read/write vector data within the a single step into two \texttt{FEEvaluation} objects, which are part of the \texttt{Matrix-free} framework in \texttt{deal.II}.
However this had no observable (positive) influence on the performance, neither for quadratic nor for quartic FEs.
Most likely because the second part of the row is already in the cache thanks to row-major layout and alignment of each dense row to cache boundaries.

\subsection{Comparison between Intel Xeon Ivy Bridge and Intel Xeon Cascade Lake}

\begin{figure}[!ht]
    \begin{subfigure}[b]{0.49\textwidth}
        \centering
        \includegraphics[width=\textwidth]{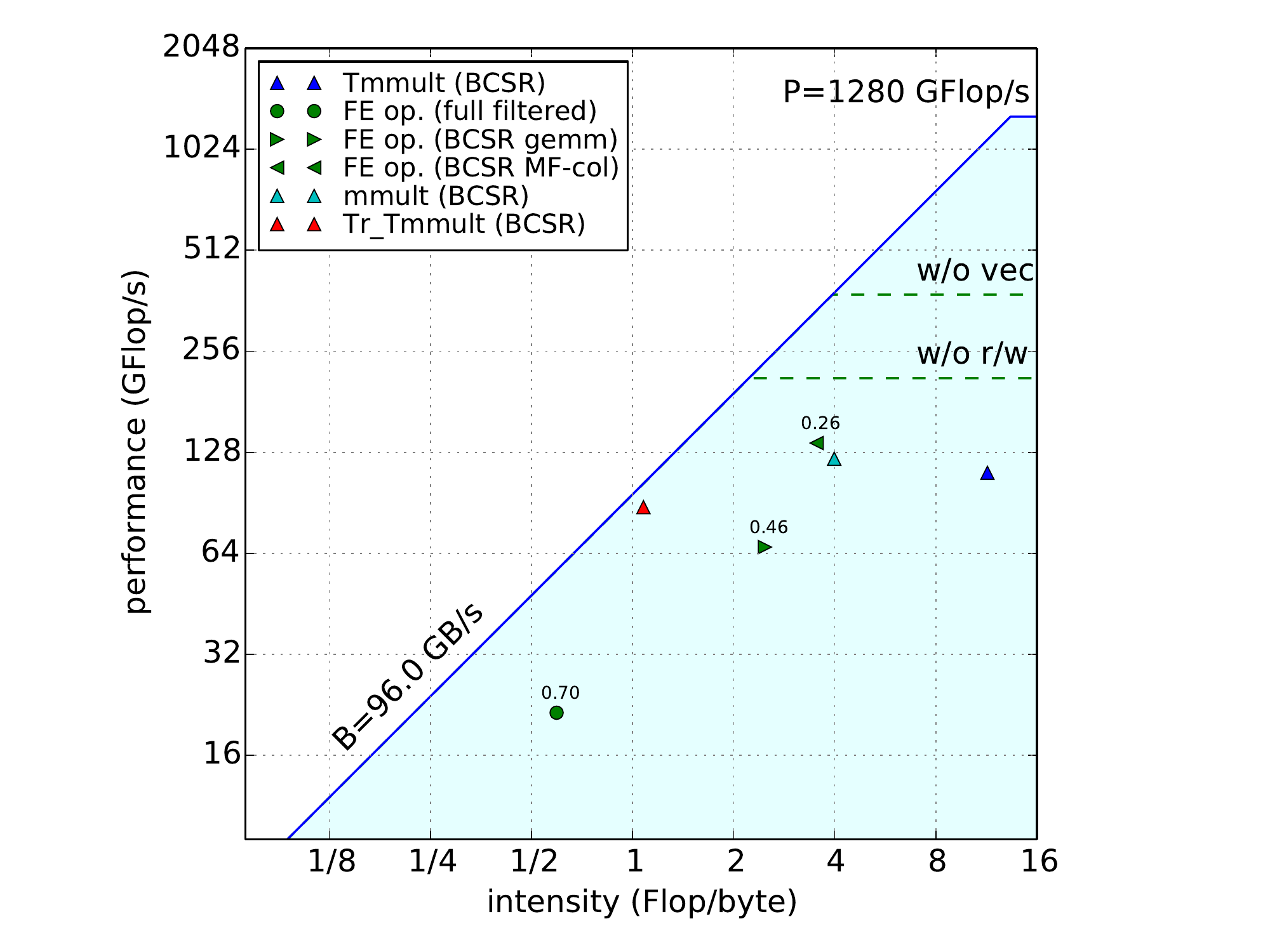}
        \caption{Quadratic elements}
        \label{fig:roofline_2csl}
    \end{subfigure}
    ~
    \begin{subfigure}[b]{0.49\textwidth}
        \centering
        \includegraphics[width=\textwidth]{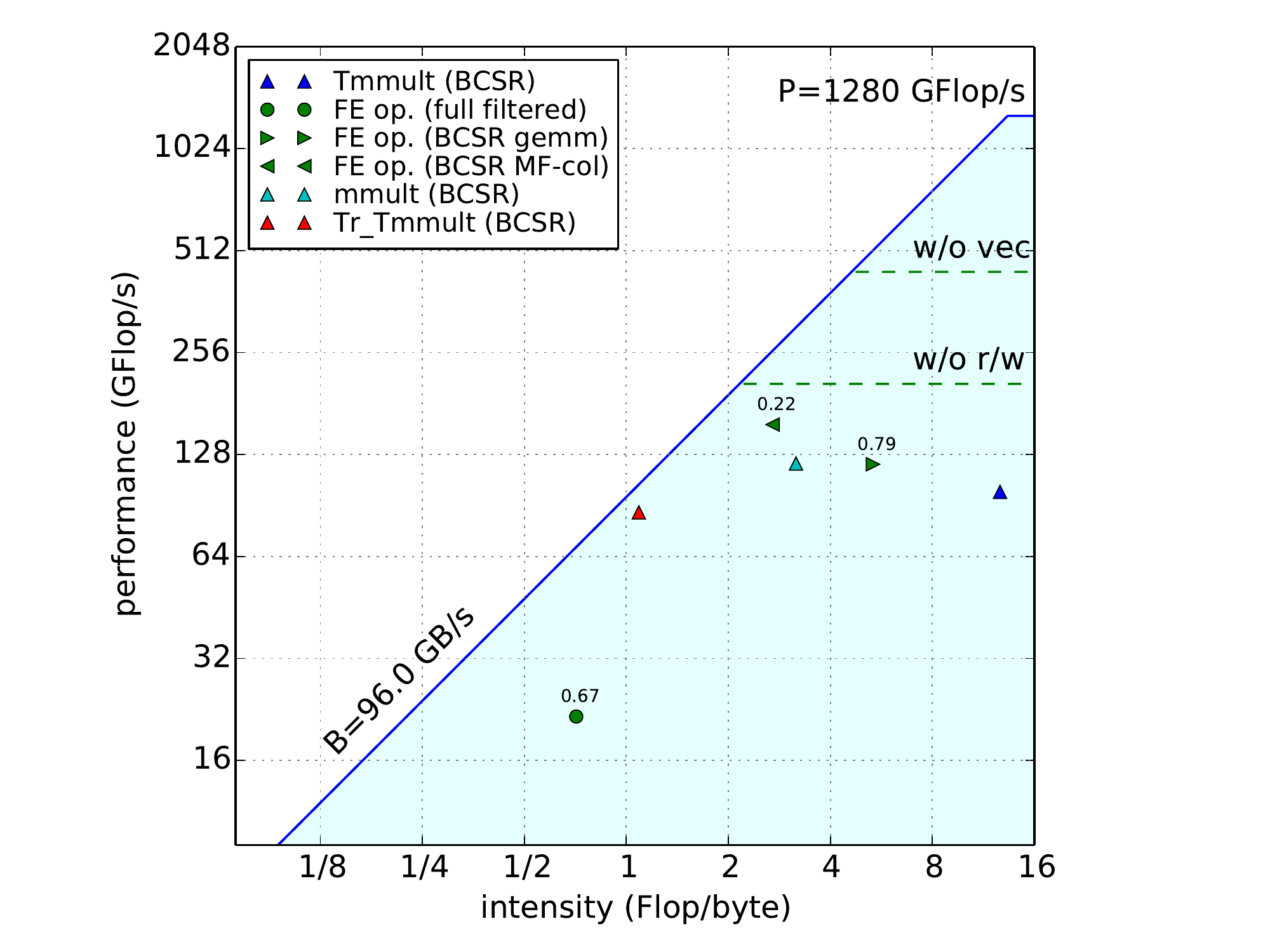}
        \caption{Quartic elements}
        \label{fig:roofline_4csl}
    \end{subfigure}
    \caption{Roofline performance model on Intel Xeon Cascade Lake using the GCC compiler.
    The dashed roof line "w/o r/w" represents the matrix-free Algorithm \ref{alg:bcsr_mf} without reading from and writing into multivectors (i.e., steps 9 and 11),
    whereas "w/o vec" represents Algorithm \ref{alg:bcsr_mf} without any multivector operations (i.e., steps 1,2,9,11 and 16). The absolute run time per call for the three matrix-free flavors is given by numbers next to the respective symbols.
    }
    \label{fig:roofline_csl}
\end{figure}

\begin{table}
    \centering
    \begin{tabular}{|r|c|c|}
    \hline
                    & quadratic & quartic \\
    \hline
    DoFs per cell   & \datacslptwodofspercell & \datacslpfourdofspercell \\
    number of cells & \datacslptwoncells      &  \datacslpfourncells \\
    DoFs/rows       & \datacslptwodofs        & \datacslpfourdofs \\
    columns         & \datacslptwocolumns     & \datacslpfourcolumns \\
    number of row blocks & \datacslptwonrowblocks & \datacslpfournrowblocks \\
    number of column blocks & \datacslptwoncolblocks & \datacslpfourncolblocks \\
    row blocks size & \pgfmathprintnumber{\datacslptworowblocksmean} $\pm$ \pgfmathprintnumber{\datacslptworowblocksstd} & \pgfmathprintnumber{\datacslpfourrowblocksmean} $\pm$ \pgfmathprintnumber{\datacslpfourrowblocksstd} \\
    column blocks size & \pgfmathprintnumber{\datacslptwocolblocksmean} $\pm$ \pgfmathprintnumber{\datacslptwocolblocksstd} & \pgfmathprintnumber{\datacslpfourcolblocksmean} $\pm$ \pgfmathprintnumber{\datacslpfourcolblocksstd} \\
    locally owned row blocks in $\overline{\gz H}$ & \pgfmathprintnumber{\datacslptwocolblocksmpimean} $\pm$ \pgfmathprintnumber{\datacslptwocolblocksmpistd} & \pgfmathprintnumber{\datacslpfourcolblocksmpimean} $\pm$ \pgfmathprintnumber{\datacslpfourcolblocksmpistd} \\
    non-empty column blocks for FE cell op. & \pgfmathprintnumber{\datacslptwomfnonemtpycolumnblocksmean} $\pm$ \pgfmathprintnumber{\datacslptwomfnonemtpycolumnblocksstd} & \pgfmathprintnumber{\datacslpfourmfnonemtpycolumnblocksmean} $\pm$ \pgfmathprintnumber{\datacslpfourmfnonemtpycolumnblocksstd} \\
    \hline
    \end{tabular}
    \caption{Statistics for the example with $320$ atoms run on 20 MPI processes.}
    \label{tab:bcsr_statistics_csl}
\end{table}

As a next experiment, we evaluate the algorithm on a newer architecture, the Intel Xeon Gold
6230 (codenamed Cascade Lake) with $20$ cores per socket. The arithmetic peak performance is 1280 GFlop/s (turbo mode enabled, maximal AVX-512 frequency is 2.0
GHz) and the peak memory bandwidth is 140 GB/s (6 channels of DDR4-2933). Since 96 GB/s
are measured for a stream copy benchmark, this number is used as a memory bandwidth
limit on this architecture.
The code is compiled with the GCC compiler version 8.1 with flags ``-O3 -march=native'', OpenMPI version 3.1.4 and Intel MKL 2018.2.
We run an example with $320$ atoms on 20 MPI
processes, which results in
very similar row and column block sizes to those considered in the previous section
(compare Table \ref{tab:bcsr_statistics_csl} and Table \ref{tab:bcsr_statistics}). Figure~\ref{fig:roofline_csl} plots the achieved performance on the Cascade Lake
architecture in terms of a roofline plot. The algorithm ``FE op. (BCSR MF-col)'' reaches
an arithmetic throughput of \pgfmathprintnumber{\datacslptwoGFlopsHBcsr} GFlop/s
for $p=2$ and \pgfmathprintnumber{\datacslpfourGFlopsHBcsr} GFlop/s for $p=4$,
as well as a memory throughput of \pgfmathprintnumber{\datacslptwoBandWidthHBcsr} GB/s
for $p=2$ and \pgfmathprintnumber{\datacslpfourBandWidthHBcsr} GB/s for $p=4$,
respectively. Compared to the Ivy Bridge architecture, the run time per core on
Cascade Lake is 40\% faster (e.g. \pgfmathprintnumber{\datacslpfourWallHBcsr}s
versus \pgfmathprintnumber{\dataGCCpfourWallHBcsr}s for $p=4$), which shows the
gains from the newer microarchitecture with AVX-512 SIMD vectorization (8-wide)
and fused multiply-add (FMA) instructions.

The roofline plot of Figure~\ref{fig:roofline_csl} shows that the two empirical
rooflines, the lower one without the read/write operations into the sparse vectors
and the higher one only involving the arithmetic work of sum factorization, are
farther away from the achieved performance. This increasing gap can be explained by the fact that the in-core performance
of the sum factorization kernels in the ``w/o vec'' part is almost four times faster
per core (or eight times faster per socket). As a consequence, the cost of data access increases in
importance and represents the main bottleneck on Cascade Lake. Especially the
memory-intensive stage of zeroing the destination vector and the MPI communication
contribute $\pgfmathadd{\datacslpfourStackMPI}{\datacslpfourStackZero}\pgfmathprintnumber[fixed]\pgfmathresult$\% of the run time for $p=4$
(comapred to $\pgfmathadd{\dataGCCpfourStackMPI}{\dataGCCpfourStackZero}\pgfmathprintnumber[fixed]\pgfmathresult$\% on Ivy Bridge with GCC compiler).
Since the arithmetically
intense local integrals of ``w/o vec'' can only be overlapped with the read/write
access but not the zeroing of the vector and the MPI exchange, the overall memory
throughput of $\pgfmathprintnumber{\datacslpfourBandWidthHBcsr}$~GB/s ends up clearly
below the theoretical value of 96~GB/s.

Taking also the algorithmic alternatives into account, the comparison between Ivy
Bridge and Cascade Lake shows that the proposed algorithm is especially attractive
on newer architectures with wider vectorization and higher Flop/Byte ratio.

\subsection{Inter-node scaling}

\begin{figure}[!ht]
    \begin{center}
    \begin{subfigure}[b]{0.49\textwidth}
        \centering
        \includegraphics[trim=0mm 0mm 0mm 0mm, width=\textwidth]{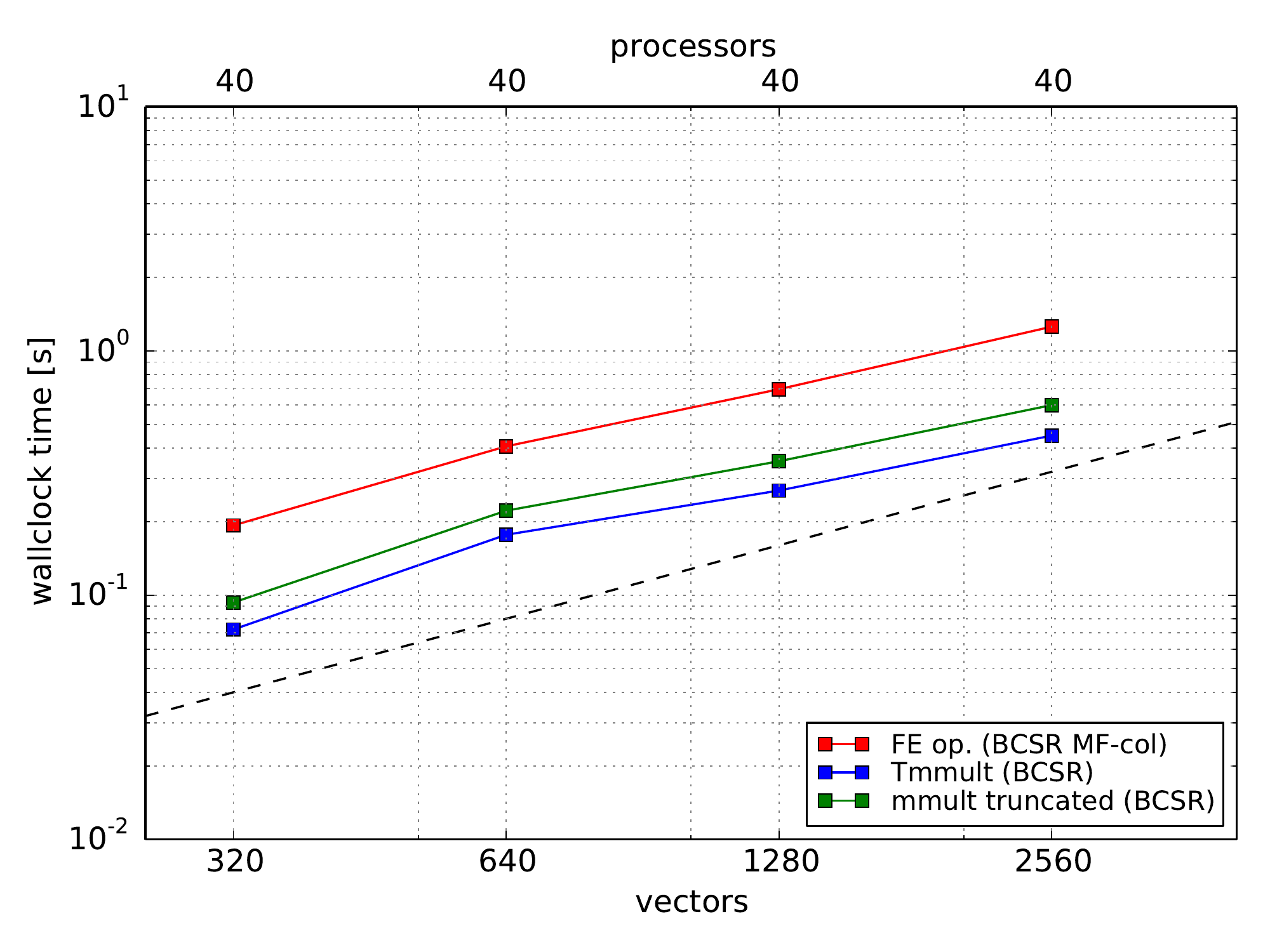}
        \caption{Quadratic elements}
        \label{fig:scaling_atoms_2}
    \end{subfigure}
    ~
    \begin{subfigure}[b]{0.49\textwidth}
        \centering
        \includegraphics[trim=0mm 0mm 0mm 0mm, width=\textwidth]{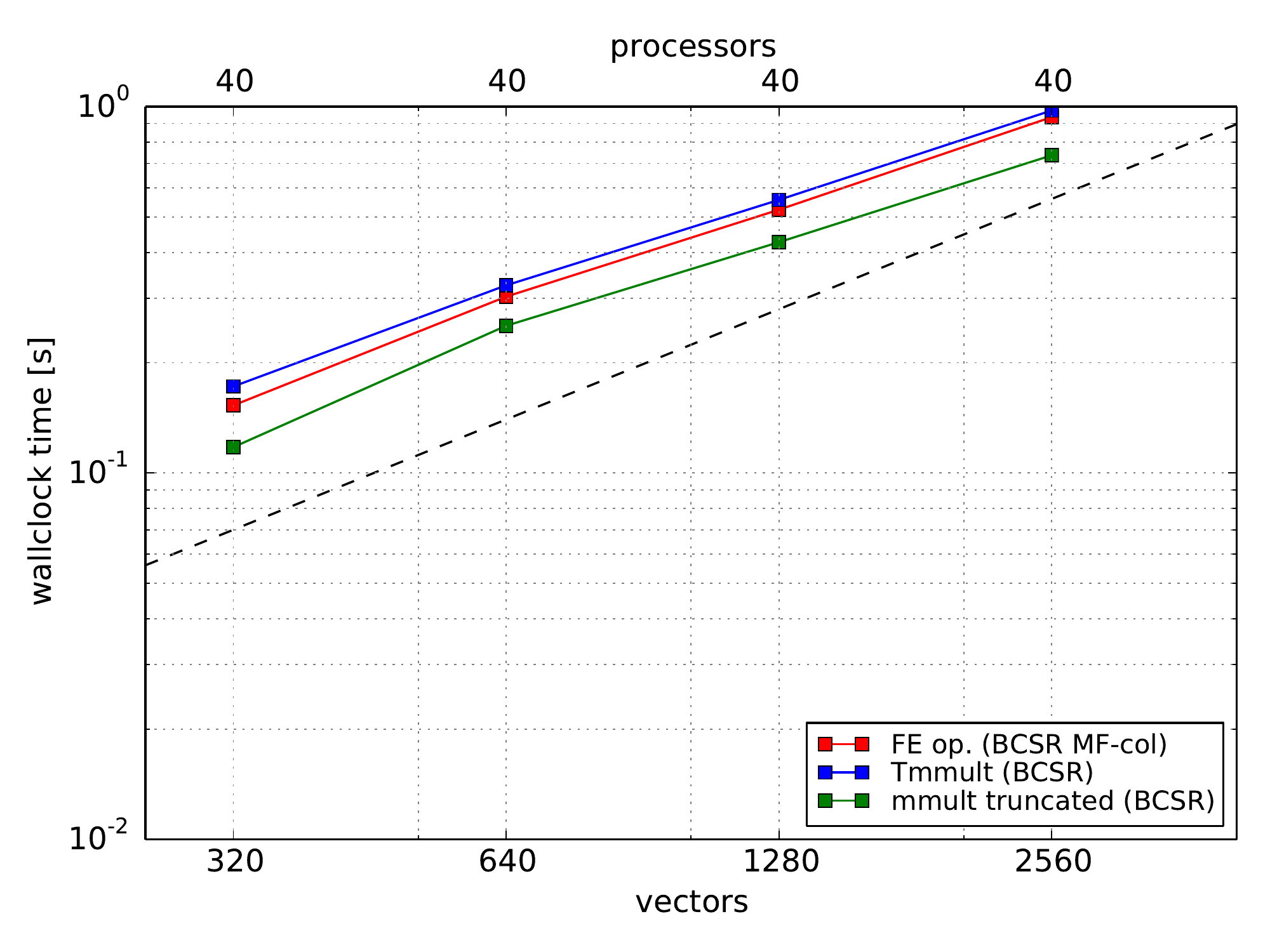}
        \caption{Quartic elements}
        \label{fig:scaling_atoms_4}
    \end{subfigure}
    \end{center}
    \caption{Scaling results for carbon nanotubes of different lengths for a fixed number of processors.
    Timings are the average of each operation called thirty times.
    }
    \label{fig:scaling_atoms}
\end{figure}

\begin{figure}[!ht]
    \begin{center}
    \begin{subfigure}[b]{0.49\textwidth}
        \centering
        \includegraphics[trim=0mm 0mm 0mm 0mm, width=\textwidth]{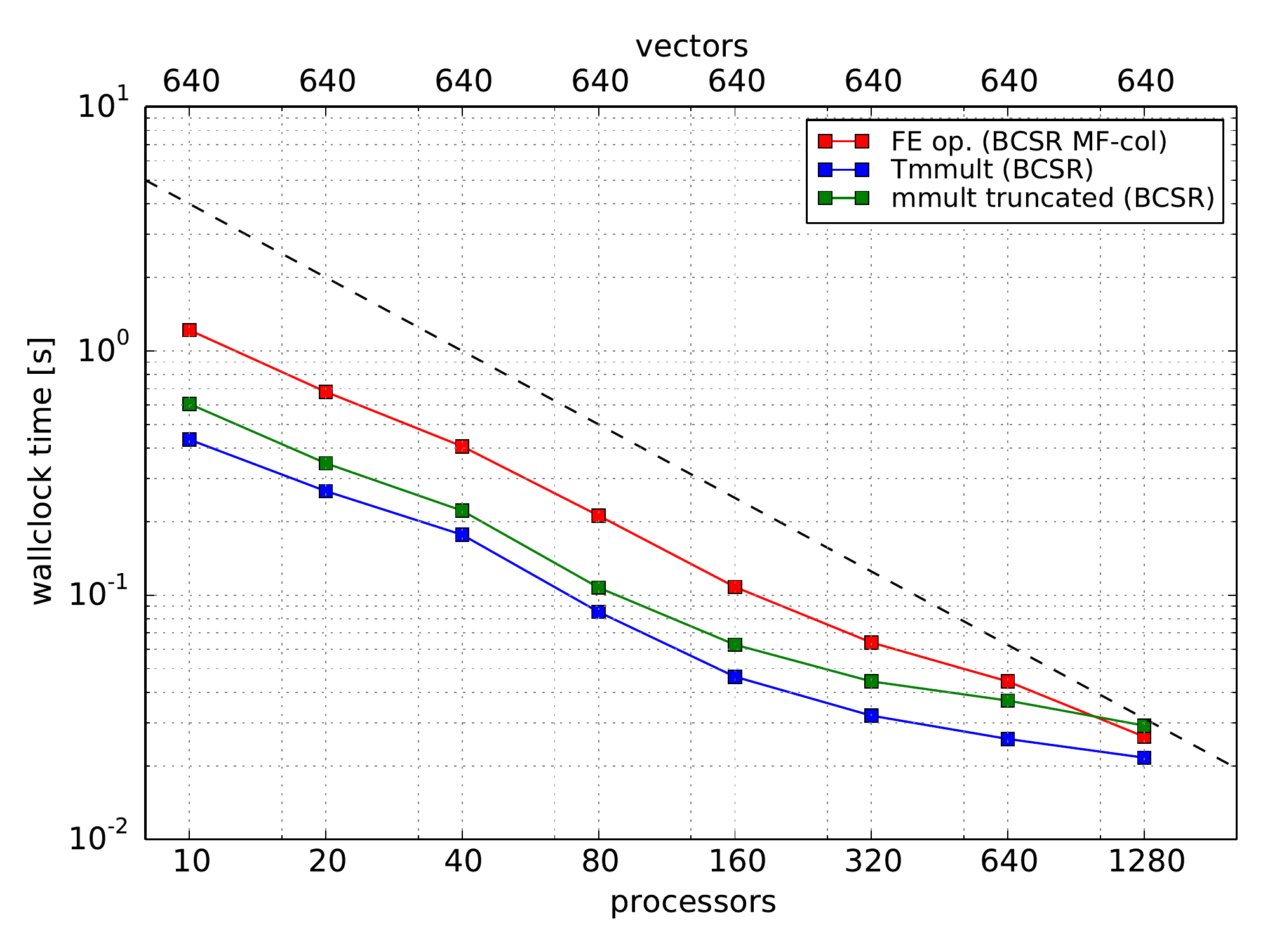}
        \caption{Quadratic elements}
        \label{fig:scaling_cores_2}
    \end{subfigure}
    ~
    \begin{subfigure}[b]{0.49\textwidth}
        \centering
        \includegraphics[trim=0mm 0mm 0mm 0mm, width=\textwidth]{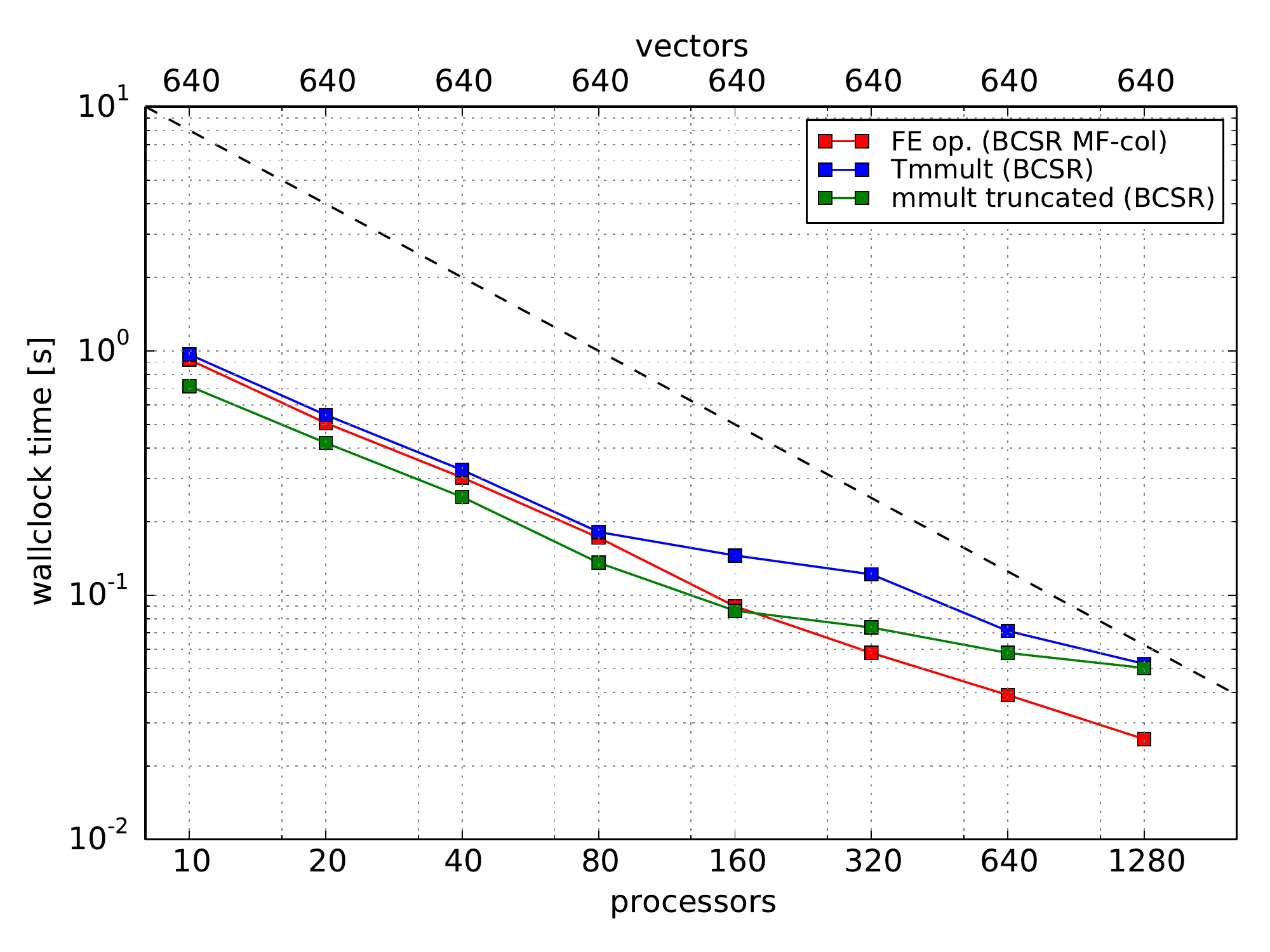}
        \caption{Quartic elements}
        \label{fig:scaling_cores_4}
    \end{subfigure}
    \end{center}
    \caption{Strong scaling results for carbon nanotube with $320$ atoms.
    Timings are the average of each operation called thirty times.
    }
    \label{fig:scaling_cores}
\end{figure}

\begin{figure}[!ht]
    \begin{center}
    \begin{subfigure}[b]{0.49\textwidth}
        \centering
        \includegraphics[trim=0mm 0mm 0mm 0mm, width=\textwidth]{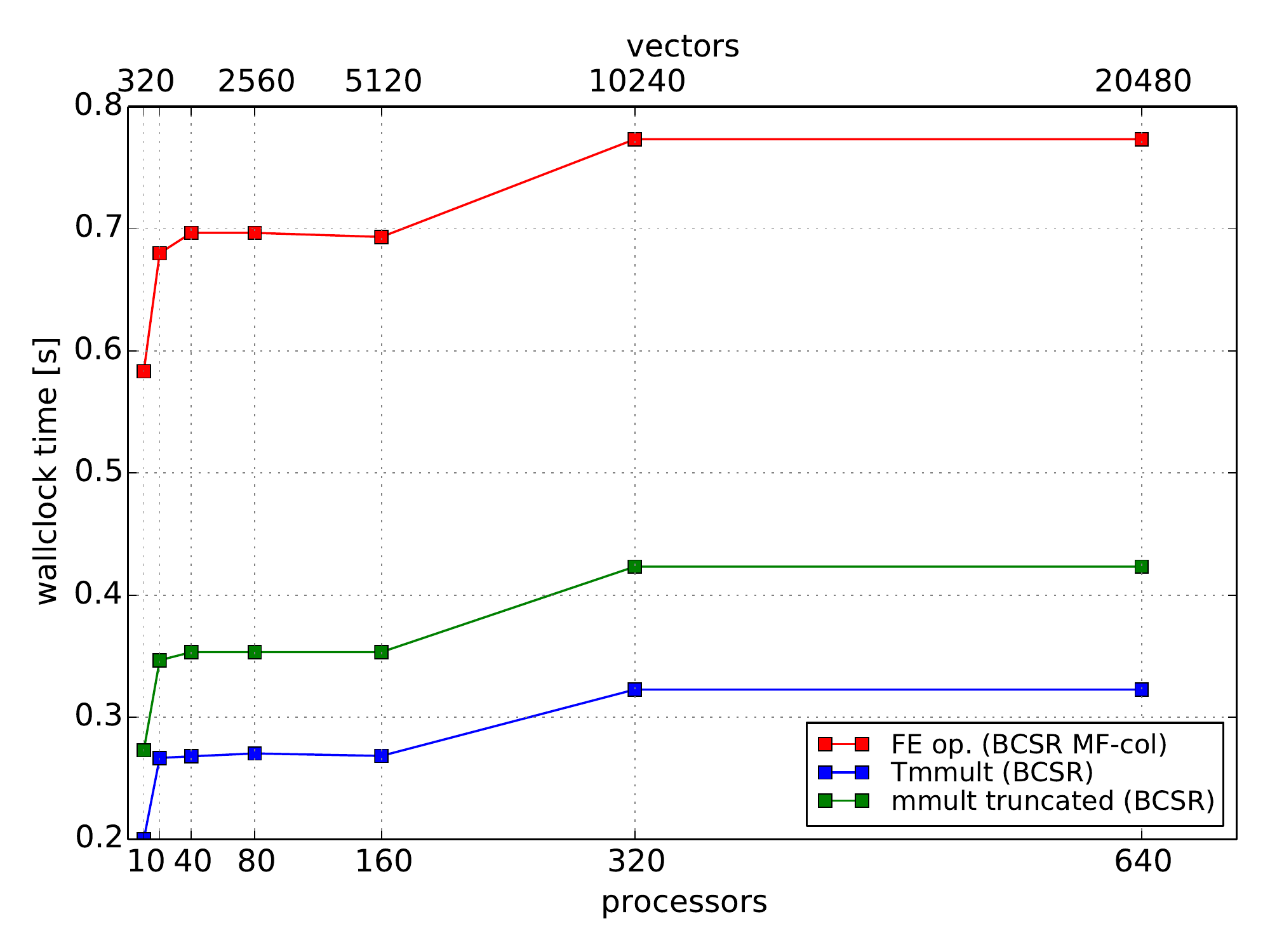}
        \caption{Quadratic elements}
        \label{fig:scaling_cores_w_2}
    \end{subfigure}
    ~
    \begin{subfigure}[b]{0.49\textwidth}
        \centering
        \includegraphics[trim=0mm 0mm 0mm 0mm, width=\textwidth]{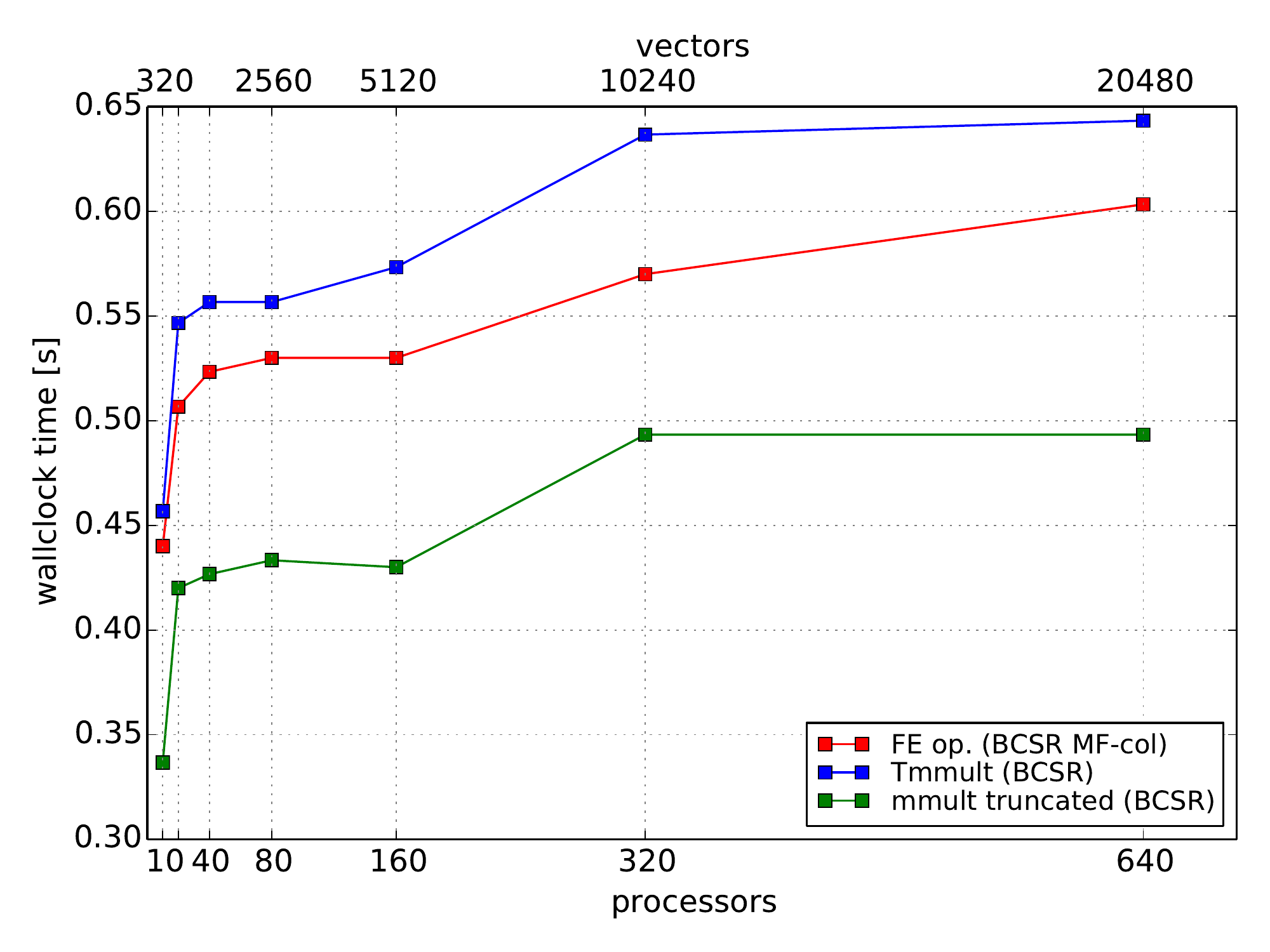}
        \caption{Quartic elements}
        \label{fig:scaling_cores_w_4}
    \end{subfigure}
    \end{center}
    \caption{Weak scaling results for carbon nanotubes of different length.
    Timings are the average of each operation called thirty times.
    }
    \label{fig:scaling_cores_w}
\end{figure}

Finally, we show results for the inter-node scaling of the proposed algorithms.
In this section, we present results for the setup with Intel compiler and Intel MPI. Even though the GCC compiler delivers slightly faster binaries on the node level,
we found the combination GCC+OpenMPI to lead to inferior scaling results.
We speculate that this is related to the fact that projected matrices (e.g. $\overline{\gz H}$) are too small,
which makes very few MPI processes own row blocks (see Tables \ref{tab:bcsr_statistics} and \ref{tab:bcsr_statistics_csl}
for examples with ten and twenty MPI processes)
and consequently results in many point-to-point MPI communication in flight during \texttt{Tmmult} ``compress'' stage, for which the Intel MPI library is more optimized than OpenMPI.
Alternative libraries such as Intel MPI or MPICH combined with GCC were not
available on the Emmy HPC cluster at the time of writing.

In our experiments, we record how the wallclock time of the required operations scales with respect to
\begin{itemize}
\item[(i)] the number of sparse vectors $M$ for a fixed number of processes $P$, see Figure \ref{fig:scaling_atoms};
\item[(ii)] the number processes $P$ for a fixed number of vectors $M$, known as strong scaling, see Figure \ref{fig:scaling_cores};
\item[(iii)] the number of processes $P$ with the number of vectors $M$ proportional to $P$, known as weak scaling, see Figure \ref{fig:scaling_cores_w}.
\end{itemize}
The results from Figures~\ref{fig:scaling_atoms} and \ref{fig:scaling_cores_w}
demonstrate an optimal algorithmic complexity with a fixed work per column vector.
Furthermore, the strong scaling results in Figure~\ref{fig:scaling_cores} illustrate
that the optimal algorithmic complexity is combined with a good distribution of
work and that no communication bottlenecks are present. The proposed column-based
matrix-free operator evaluation shows very good scalability all the way to 1280~MPI
ranks or around $2.5\cdot 10^{-2}$ seconds for a single operator evaluation on
640~vectors. This number can be compared to the scaling limit of around
$2\cdot 10^{-4}$ seconds for an operator evaluation on a single vector with
matrix-free algorithms \cite{Kronbichler2019}. Thus, the proposed algorithms with
block columns allow to overcome this limit and efficiently address large-scale
atomic systems. The parallel speedup of the \texttt{Tmmult} and \texttt{mmult}
operations is excellent for enough parallelism as well, but it levels off for
very large sizes. This is expected given the more global communication nature
along the long and skinny matrix dimensions.

Further improvements could probably be made by overlapping the MPI communication
with computations both for the matrix-free FE operator as well as the matrix-matrix products.
The matrix-free framework of \texttt{deal.II} supports ordering of cells in such a way that
each MPI process first loops through cells that only contain locally owned DoFs and thus
does not need to wait for ``ghost'' values from other MPI processes.
In order to support such a setup,
the MPI communication (i.e., ``update ghost values'' and ``compress'') in the BCSR matrix structure
has been implemented as a two-stage procedure,
that involves initiation of the non-blocking point-to-point communication and waiting for it to complete.
This shall allow us to overlap computation and communication in FE operator through \texttt{deal.II}'s matrix-free framework.
We deem it possible to also hide MPI communication for matrix-matrix products
in Algorithm \ref{alg:bcsr_mmult}.
To that end BCSR row iterators could be extended so that one
can iterate only over locally owned or ghost column blocks.
Note that this is similar to the approach adopted in PETSc \cite{petsc-user-ref}
for matrix-matrix multiplication,
where each MPI process stores the locally owned part of a sparse matrix as two blocks:
one square diagonal block with column indices corresponding to locally owned rows in the domain vector,
and one block which stores all the other columns corresponding to ghosted entries.
We leave investigation of these (and possibly other) aspects of MPI scaling to further studies.
The results of this section should rather be taken as a proof of concept, that the BCSR storage format with 1D row-wise MPI partitioning is a promissing approach for the required operations on tall and skinny sparse multivectors with the FE matrix-free operators.

\section{Summary}
\label{sec:summary}

In this contribution we have proposed algorithms for sparse finite element multivectors suitable for matrix-free implementations
using BCSR matrices with row-wise partitioning to utilize MPI and SIMD vectorization over columns.
In addition to standard matrix-matrix products, we developed tailored algorithms and data structures to efficiently support the matrix-free evaluation of
operators in the sparse vector context.
To that end we proposed a way to integrate ghost DoFs into the BCSR structure and implemented the required MPI non-blocking communication.
Our single-node performance studies demonstrate that the loop overhead due to BCSR format consumes less than 5\% of the walltime, while enabling the study of big atomic systems in linear time.
For higher polynomial degrees the BCSR matrix-free operator outperforms both the cell matrix based \texttt{dgemm} variant as well as column vectors based matrix-free counterparts.
We were able to achieve around on fourth of the maximum arithmetic performance of a 10-core Intel Ivy Bridge processor and around an eighth of the arithmetic peak on a 20-core Intel Cascade Lake processor. The latter also saturates up to 60\% of the available memory bandwidth. We have shown that the performance gap can be explained by the cost of data access in the matrix-free operator evaluation, such as the MPI communication or as simple operations as zeroing the destination vector. The BCSR-specific components within the matrix-free operator have been shown to only marginally increasing the run time, showing that the proposed algorithms are efficient.
The inter-node scaling results confirm that
the advocated BCSR storage format is a good candidate for efficient and scalable
operations on sparse multivectors within the context of the FE method applied to problems in quantum mechanics.

In order to enable the large-scale finite element solution of DFT via the orbital
minimization approach with BCSR multivectors, a few additional steps are necessary,
such as the extension of the proposed algorithms for adaptive mesh refinement for
which ideas from \cite{Kronbichler2012} can be used, a more transparent handling
of column-based evaluation in terms of the metric terms within the \texttt{deal.II}
library, as well as geometric multigrid preconditioners to improve the convergence
of minimizers \cite{Davydov2018}. For the latter, efficient intergrid matrix-free
transfer operators would need to be developed.
This will be the focus of our future work in this direction. Eventually,
performance of the proposed matrix-free kernels for BCSR setups could be improved
by limiting message-passing communication of ghosts only between the nodes, and
use shared memory concepts within the nodes.

\begin{acks}

DD is grateful to Georg Haager (FAU), Jan Eitzinger(FAU) and Thomas Gruber (FAU)
for fruitful discussions on node level performance measurements using LIKWID tool.
DD acknowledge the financial support of the German Research Foundation (Deutsche Forschungsgemeinschaft, DFG), grant DA 1664/2-1
and
the Bayerisches Kompetenznetzwerk
f\"ur Technisch-Wissenschaftliches Hoch- und H\"ochstleistungsrechnen
(KONWIHR).
MK was partly supported by  the  German Research Foundation (DFG) under the project
"High-order  discontinuous Galerkin for the EXA-scale" (ExaDG) within the priority program
"Software for Exascale Computing" (SPPEXA), grant agreement no. KR4661/2-1.

\end{acks}

\bibliographystyle{ACM-Reference-Format}
\bibliography{bibliography}

\end{document}